\documentclass[aps,pre,twocolumn,superscriptaddress,floatfix]{revtex4-1}
\usepackage{amsmath,amssymb}
\usepackage{bbm}
\usepackage{graphicx} 
\usepackage{color}

\newcommand{\IP}{{\bf I}^{\rm (PP)}}
\newcommand{\IA}{{\bf I}^{\rm (AA)}}

\newcommand{\NP}{N^{\rm (P)}}
\newcommand{\NA}{N^{\rm (A)}}
\newcommand{\myNG}{N^{\rm (G)}}
\newcommand{\xiP}{\xi^{\rm (P)}}
\newcommand{\xiA}{\xi^{\rm (A)}}

\newcommand{\KP}{{\bf K}^{\rm (PP)}}
\newcommand{\KA}{{\bf K}^{\rm (AA)}}

\newcommand{\JPP}{{\bf J}^{\rm (PP)}}
\newcommand{\JAA}{{\bf J}^{\rm (AA)}}
\newcommand{\JPA}{{\bf J}^{\rm (PA)}}
\newcommand{\JAP}{{\bf J}^{\rm (AP)}}

\newcommand{\APA}{{\bf A}^{\rm (PA)}}
\newcommand{\AAP}{{\bf A}^{\rm (AP)}}

\newcommand{\cS}{\mathcal{S}}

\newcommand{\bbB}{\mathbb{B}}
\newcommand{\bbI}{\mathbb{I}}
\newcommand{\bbJ}{\mathbb{J}}
\newcommand{\bbA}{\mathbb{A}}

\newcommand{\bbK}{\mathbb{K}}

\newcommand{\bbH}{\mathbb{H}}

\newcommand{\tm}{\tilde{m}}

\newcommand{\bk}{\bar{k}}
\newcommand{\tr}{\tilde{r}}

\newcommand{\tLambda}{\tilde{\Lambda}}

\newcommand{\tA}{\tilde{A}}
\newcommand{\tx}{\tilde{x}}
\newcommand{\tcP}{\tilde{c}^{\rm (P)}}
\newcommand{\tcA}{\tilde{c}^{\rm (A)}}
\newcommand{\tcG}{\tilde{c}^{\rm (G)}}

\newcommand{\bbtc}{\mathbbm{\tilde{c}}}

\newcommand{\bbtJ}{\mathbb{\widetilde{J}}}
\newcommand{\bbtK}{\mathbb{\widetilde{K}}}
\newcommand{\bbtA}{\mathbb{\widetilde{A}}}
\newcommand{\bbtB}{\mathbb{\widetilde{B}}}
\newcommand{\bbtN}{\mathbb{\widetilde{N}}}

\begin{document}
\title{Stability and selective extinction in complex mutualistic networks}
\author{Hyun Woo Lee}
\author{Jae Woo Lee}
\email{jaewlee@inha.ac.kr}
\affiliation{Department of Physics, Inha University, Incheon 22212, Korea}
\author{Deok-Sun Lee} 
\email{deoksunlee@kias.re.kr}
\affiliation{School of Computational Sciences and Center for AI and Natural Sciences, Korea Institute for Advanced Study, Seoul 02455, Korea}
\date{\today}

\begin{abstract}
We study species abundance in the empirical plant-pollinator mutualistic networks exhibiting broad degree distributions, with uniform intra-group competition assumed, by the Lotka-Volterra equation.  The stability of a fixed point is found to be identified by the signs of its non-zero components and those of its neighboring fixed points. Taking the annealed approximation, we derive the non-zero components  to be formulated in terms of degrees and the rescaled interaction strengths, which lead us to find different stable fixed points depending on parameters, and we obtain the phase diagram. The selective extinction phase finds small-degree species extinct and effective interaction reduced, maintaining stability and hindering the onset of instability. The non-zero minimum species abundances from different empirical networks show data collapse when rescaled as predicted theoretically.
 \end{abstract}
\maketitle

\section{Introduction}
\label{sec:intro}

A community of randomly interacting species can become unstable as the number of species and their interaction connectivity together go beyond a threshold~\cite{MAY:1972aa}.  Such a random-interaction model can be informative with the help of the random matrix theory, and it has been instrumental in the theoretical study of ecological systems, illuminating their features from the perspective of stability~\cite{Goh1979,Allesina2015,Stone2016,Grilli:2017wj,PhysRevE.95.042414,PhysRevLett.125.048101,doi:10.1098/rsif.2019.0391,*Pettersson2020}. Recently available data-sets point out the complex organization of interspecific interactions,  neither completely random nor ordered~\cite{MONTOYA2002405,Dunne2002,Bascompte2003,Jordano2003,Montoya2006,Bascompte2007,Guimaraes2007, Olesen2007,thebault10,Maeng2011}, and they have drawn the attention of researchers to the origins and implications of  over-represented network structural features~\cite{Bastolla2009,suweis2013,https://doi.org/10.1002/ece3.1930,GanYan2017}.

Contrary to the unstructured communities in which every species is subject to the identical randomness in its interaction profile, individual species can be in fundamentally different states under structured interactions. For instance,  the mutualistic partnership between flowering plants and pollinating bees is  characterized by different numbers of partners, called degrees, from species to species.  Considering also the  intrinsic competitions among plants and among pollinators due to limited resources~\cite{Maeng2012,suweis2013,Pascual-Garcia:2017wj,barbier_generic_2018,Gracia-Lazaro:2018tg,wang2021interspecific,Maeng2019,Cai:2020ta}, one finds that the abundance of a species increase due to the benefit from mutualistic partners but also decrease due to the cost from competition, the imbalance of which may  lead some species to flourish but others to become extinct~\cite{Gracia-Lazaro:2018tg,wang2021interspecific}. The mechanism driving such different fates across species  remains to be elucidated~\cite{James:2012aa,Saavedra:2013aa,Allesina2012,Stone2020}.

Here we investigate the different abundances and different likelihood of extinction of individual species  in heterogeneous mutualistic networks from the perspective of stability. We consider the Lotka-Volterra-type (LV) equation for species abundance on plant-animal mutualistic networks, with the mutualistic interaction constructed from an empirical dataset~\cite{web_of_life}, which is heterogeneous, and all-to-all intra-group competition assumed~\cite{Maeng2012,suweis2013,Stone2016,Pascual-Garcia:2017wj, barbier_generic_2018,Gracia-Lazaro:2018tg,wang2021interspecific,Maeng2019,Cai:2020ta}. The strengths of mutualism and competition are set to be uniform. We restrict ourselves to the stationary state and study the stable fixed points. Exponentially many fixed points exist with zero components at different species, but only the stable one is relevant to the stationary state. 

To find the stable fixed point,  we first show that stability can be assessed by  the signs of the non-zero components of the considered fixed point and  its neighboring fixed points. Next, approximating the adjacency matrix to be in factorized form, we derive the non-zero components of each fixed point to be formulated in terms of degrees  and the rescaled interaction strengths. Using these results, we devise an algorithm to classify species into surviving and extinct ones and thereby formulate the stable fixed point, which turns out to work well as supported by  good agreement with numerical solutions. The  extinction or the diverging abundances of selected species happens depending on parameters, the analytic understanding of which allows us to obtain the phase diagram, including the full coexistence, selective extinction, and unstable phase. In the selective extinction phase, small-degree species go extinct, which results in reducing the effective interaction among the surviving species and suppressing the onset of instability. Our study enables a principled discrimination between surviving and extinct species and the prediction of  the abundances of the surviving species, helping us to understand the interplay of stability, species abundance, and extinction in structured ecological communities.

\section{Model}
\label{sec:model}

We consider a system of  $\NP$ flowering plant species and  $\NA$ pollinating  animal  species. Their abundances $x_i$'s evolve with time under the LV equation  as 
\begin{equation}
{dx_i(t) \over dt}
 = x_i \left(\alpha_i + \sum_{j=1}^S B_{ij} x_j\right), 
\label{eq:LV}
\end{equation}
where $S=\NP+\NA$ is the total number of species, $\alpha_i=1$ is the intrinsic growth rate, and $\bbB=(B_{ij})$ is the  $S\times S$  interaction matrix  
\begin{equation}
\bbB \equiv  -\mathbb{I}-c(\mathbb{J}^{(0)} - \mathbb{I}) + m \mathbb{A}.
\label{eq:B}
\end{equation}
Here  $\bbI\equiv\begin{pmatrix} \IP & {\bf 0} \\ {\bf 0} & \IA\end{pmatrix} = \IP \oplus \IA $ is the identify matrix representing intraspecific regulation, $\bbJ^{(0)} \equiv \begin{pmatrix} \JPP & {\bf 0} \\ {\bf 0} & \JAA\end{pmatrix} =\JPP \oplus \JAA$ consists of  the matrices of $1$'s ($J_{pp^\prime}=J_{aa^\prime}=1$ for all $p, p^\prime, a, a^\prime$) representing all-to-all competition among plants  and among pollinators along with strength $0<c<1$, and $\bbA\equiv \begin{pmatrix} {\bf 0} & \APA \\ \AAP & {\bf 0} \end{pmatrix} =\APA \oplus \AAP$ is the symmetric adjacency matrix ($A_{pa}=A_{ap}=0, 1$) with $\AAP = {\APA}^\intercal$ representing the mutualistic interaction  along with the mutualism strength $m>0$. The useful properties of the matrices of $1$'s, which are given in  Appendix~\ref{sec:Jmat}, enable various analytic calculations. There are $L\equiv \sum_{p,a} A_{pa}$ mutualistic partner pairs.  
 
\begin{figure}
\centering
\includegraphics[width=\columnwidth]{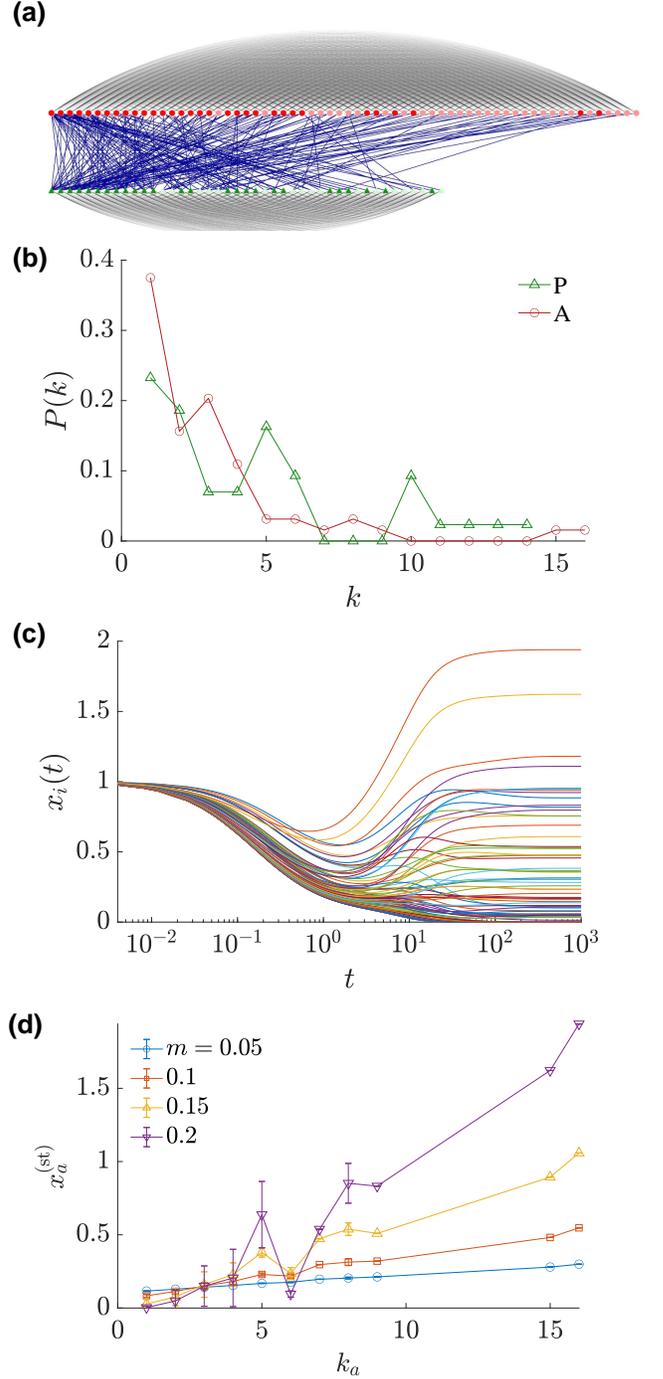}
\caption{Interaction network and species abundance of the selected community.  (a) Interaction network of $\NP=43$ plants species (green triangle) and $\NA=64$  animal species (red circle) connected by $L=196$ mutualistic interaction links (blue) and all-to-all intra-group competition links (gray). Nodes of light red and light green represent extinct species for $c=0.1$ and $m=0.2$. 
(b) Mutualism degree distributions for plants (triangle) and  animals (circle).
(c) Abundances of individual species  (different lines) for $c=0.1$ and $m=0.2$.
(d) The stationary-state abundance vs.  degree for animal species  with $c=0.1$ and selected $m$'s.}
\label{fig:model}
\end{figure}

We select a real-world community 
~\cite{https://doi.org/10.1002/j.1537-2197.1982.tb13237.x} in a database~\cite{web_of_life}  to construct the adjacency matrix $\bbA$ and use it to define $\bbB$ by Eq.~(\ref{eq:B}), and build all our theoretical framework, which will be applied later to other communities. Notice that the elements of $\bbB$ are not  random numbers but represent the interaction relationships among different species with a uniform interaction strength $c$ or $m$. The whole interaction network encoded in $\bbB$ and the distributions of the mutualism degrees, $k_p \equiv \sum_a A_{pa}$ of plants  and $k_a\equiv\sum_p A_{pa}$ of animals, are presented in Figs.~\ref{fig:model}(a) and ~\ref{fig:model}(b), respectively. Different  degrees of species are a fundamental heterogeneity in their mutualistic interaction profiles, which have been neglected in the random-interaction model assuming the interaction strength between each pair of species  to be an independent and identically distributed random variable~\cite{MAY:1972aa,Goh1979,Allesina2015,Stone2016,Grilli:2017wj,PhysRevE.95.042414,PhysRevLett.125.048101,doi:10.1098/rsif.2019.0391,*Pettersson2020}, but  they are of main concern in the present study.

Integrating Eq.~(\ref{eq:LV}) up to $T=10^3$ with the initial condition $x_i(0)=1$, we find, as shown in Fig.~\ref{fig:model} (c), that $x_i(t)$'s for different species $i$ exhibit different behaviors as functions of time. They become stationary in the long-time limit and we approximate the stationary-state abundance by the species abundance at the final time step $T$, 
\begin{equation}
x_i^{\rm (st)}\equiv \lim_{t\to\infty} x_i(t)\simeq  x_i(T).
\label{eq:xist}
\end{equation}
Some species show $x_i^{\rm (st)}=0$, implying their extinction. Also, as shown in Fig.~\ref{fig:model} (d),  $x_i^{\rm (st)}$  tends to grow with degree $k_i$. This correlation will be clarified  in the next sections. Another remarkable feature  is that the effect of mutualism on the species abundance can be drastically different depending on species~\cite{Gracia-Lazaro:2018tg,Cai:2020ta,wang2021interspecific}; The species with large degrees find their abundances increasing with $m$ but the abundances of the species having small degrees decrease with $m$ [Fig.~\ref{fig:model} (d)]. In the next sections, we develop the analytic approach to understand the nature and origin of such heterogeneity in the species abundance depending on parameters.

\section{Stability} 
\label{sec:stability}

The state of a dynamical system, like Eq.~(\ref{eq:LV}), is expected to converge to a stable fixed point  in the long-time limit.  If a stable fixed point exists and is unique, its components will give the stationary-state abundances  that we obtain numerically.

Depending on which components are zero, there are $2^S$ different fixed points of Eq.~(\ref{eq:LV}); A fixed point $\vec{x}^*=(x_i^*)$ has components 
\begin{equation}
x_i^*= 
\begin{cases} 
0 &   \  {\rm for} \ i\in \cS^{(0)}, \\
- \sum_{j\in \cS^{(+)}} ((B^{(+)})^{-1})_{ij} &  \ {\rm for} \ i\in \cS^{(+)},
\end{cases}
\label{eq:xi*}
\end{equation}
where $\cS^{(0)}$ and $\cS^{(+)}$ are the set of the species with zero and non-zero components in the considered fixed point, respectively, and $((B^{(+)})^{-1})_{ij}$ is the inverse of the effective interaction matrix $\bbB^{(+)}$ obtained by eliminating  the rows and columns of the species of $\cS^{(0)}$ in $\bbB$~\cite{doi:10.1098/rsif.2019.0391,*Pettersson2020}.   We keep the indices $i$ or $j$ of the original interaction matrix $\bbB$ such that $B^{(+)}_{ij} = B_{ij}$ as long as $i,j\in \cS^{(+)}$.

The fixed point in Eq.~(\ref{eq:xi*}) is stable if a small perturbation $\delta x_i = x_i - x_i^*$ does not grow persistently with time but vanishes in the long-time limit.  The time-evolution of the perturbation is given by  ${d\over dt} \delta x_i = H_{ij} \delta x_j$, which involves  the Jacobian matrix at the fixed point $\vec{x}^* = (x_i^*)$ given by 
\begin{equation}
H_{ij} = \delta_{ij} \left[1 + \sum_{\ell=1}^S B_{i\ell} x_\ell^*\right] + x_i^* B_{ij}.
\label{eq:Hij}
\end{equation}
For $i\in \cS^{(0)}$, it holds that $x_i^*=0$, and therefore one can see that $H_{ij}=\delta_{ij} \left[1 + \sum_{\ell=1}^S B_{i\ell} x_\ell^*\right]$. For  $i\in \cS^{(+)}$, it holds that $1+\sum_\ell B_{i\ell} x_\ell^* = 1+\sum_{\ell\in \cS^{(+)}} B^{(+)}_{i\ell} x_{\ell}^*=0$, leading to $H_{ij}=x_i^* B_{ij}$. If all the eigenvalues $\lambda_i$'s  of $\bbH=(H_{ij})$ have  negative real parts, then the small perturbation will die out and the fixed point can be considered as stable. 

We derive the approximate expression for  the eigenvalues of $\bbH$. For $i\in \cS^{(0)}$, let us consider a neighboring fixed point $\vec{x}^{*\prime} = (x_\ell^{*\prime})$ with $\cS^{(+)\prime}=\cS^{(+)} \cup \{i\}$, which satisfies $1+\sum_\ell B_{i\ell} x_\ell^{*\prime} = 1 + \sum_{\ell \in \cS^{(+)}} B_{i\ell} x_\ell^{*\prime} + B_{ii} x_{i}^{*\prime}=0$.  Assuming that $x_\ell^{*\prime} \simeq x_\ell^*$ for $\ell\in \cS^{(+)}$ and using $B_{ii}=-1$, we find $H_{ij} \simeq \delta_{ij} x_i^{*\prime}$. Therefore, with rows and columns rearranged, the Jacobian matrix $\bbH$ contains one zero submatrix, say, $H^{(0+)}_{ij}=0$, and  three non-zero block submatrices $H^{(00)}_{ij}\simeq x_i^{*\prime} \delta_{ij},  H^{(+0)}_{ij} = x_i^* B_{ij}$, and $H^{(++)}_{ij} = x_i^* B^{(+)}_{ij}$ such that 
\begin{equation}
\bbH = \begin{pmatrix}
{\bf H}^{(00)} & {\bf 0} \\
{\bf H}^{(+0)} & {\bf H}^{(++)}
\end{pmatrix} 
= \begin{pmatrix}
(\delta_{ij} x_i^{*\prime}) & {\bf 0} \\
(x_i^* B_{ij}) & (x_i^* B^{(+)}_{ij})
\end{pmatrix}.
\label{eq:Jacobian}
\end{equation}
 Given the zero block submatrix, all the eigenvalues of $\bbH$ come from the diagonal blocks, $H^{(00)}_{ij}\simeq x_i^{*\prime} \delta_{ij}$ and $H^{(++)}_{ij} = x_i^* B^{(+)}_{ij}$.   ${\bf H}^{(00)}$ is already diagonalized, with $x_i^{*\prime}$'s as its eigenvalues. To obtain the  eigenvalues of ${\bf H}^{(++)}$, we decompose it as $H^{(++)}_{ij} = -x_i^*\delta_{ij} + V_{ij}$ with  $V_{ij}\equiv - c x_i^* (1 - \delta_{ij}) + m x_i^* A_{ij}$ and apply the perturbation theory with $V_{ij}$ taken as a perturbation to obtain the approximate eigenvalues $-x_i^*$'s for small $c$ and $m$ as described in Appendix~\ref{sec:eigen} and Ref.~\cite{Stone2020}. Therefore we find that the eigenvalues $\lambda_i$'s of $\bbH$ are approximately 
 \begin{equation}
 \lambda_i \simeq 
 \begin{cases}
 x_i^{*\prime} & \ {\rm for} \ i\in \cS^{(0)},\\
 -x_i^* & \ {\rm for} \ i\in \cS^{(+)}.
 \end{cases}
 \label{eq:lambda}
 \end{equation}
A concrete example of constructing the Jacobian matrix and deriving Eq.~(\ref{eq:lambda})  for a small community is presented in Appendix~\ref{sec:eigen}. 

Using Eq.~(\ref{eq:lambda}), we can see that all the eigenvalues are negative and the fixed point in Eq.~(\ref{eq:xi*}) is stable when the following conditions are met: i) every species $i$ that would have a negative fixed-point abundance ($x_i^{*\prime}<0$) if it were added to $\cS^{(+)}$ has zero abundance  ($x_i^*=0$) and is in $\cS^{(0)}$, and ii) every species $i$  in $\cS^{(+)}$ has a positive fixed-point abundance ($x_i^*>0$).  If the components of a fixed point are known, one can use these stability conditions  to predict  which  species go extinct and which species survive. In the next section, we derive the approximate analytic formula for the components of a fixed point and use it along with Eq.~(\ref{eq:lambda}) to infer the stable fixed point.

\section{Analytic approaches to the fixed point} 
\label{sec:formula}

In this section we assume that all species survive, $\cS^{(0)}=\emptyset$, and obtain the components of the corresponding fixed point in Eq.~(\ref{eq:xi*}).  While the inverse of a matrix is not available in a closed form in general, here we first consider the case of zero or weak mutualism and then take an approximation for the adjacency matrix to derive the components of the fixed point. The obtained results will be generalized straightforwardly to the case of $\cS^{(0)}\neq\emptyset$ in Sec.~\ref{sec:sephase}.  

\subsection{No mutualism}

Let us consider the case of no mutualism but competition only. The interaction matrix for $m=0$ is given in a simple form as   
\begin{equation}
\bbB =  \bbB_0 \equiv -(1-c)\bbI - c\bbJ^{\rm(0)}.
\label{eq:B0}
\end{equation}
Trying $\bbB_0^{-1} =-{1\over 1-c}(\IP \oplus \IA \ominus b^{\rm (P)}\JPP \ominus  b^{\rm (A)}\JAA)$ as its inverse and inserting it into $\bbB_0 \bbB_0^{-1}=\bbI$, we find that $b^{\rm (P)} = {\tcP \over \NP}$ and $b^{\rm (A)} = {\tcA \over \NA}$, where  the rescaled competition strengths $\tcP$ and $\tcA$ are defined as 
\begin{equation}
\tcG \equiv {c \myNG \over c\myNG + 1-c}
\end{equation}
with G representing either P or A. The properties of the matrix of $1$'s , such as $\JPP\JPP = \NP \JPP$, are used for derivation. The rescaled competition $\tcG$ ranges between $0$ and $1$, and grows with $c$ and $\myNG$ as long as $0<c<1$. 

The inverse matrix $\bbB_0^{-1}$ is represented in a compact form as
\begin{equation}
 \bbB_0^{-1} = (1-c)^{-1}\left(-\bbI +\bbtc\bbtJ^{(0)}\right)
 \label{eq:Binv0}
 \end{equation}
with $\bbtc \equiv \tcP \IP \oplus \tcA  \IA$ and $\bbtJ^{(0)}\equiv {\JPP \over \NP}  \oplus {\JAA \over \NA}$. Then the component of the fixed point in Eq.~(\ref{eq:xi*}) with $\cS^{(0)}=\emptyset$ is given by 
\begin{equation}
x_i^{*(0)} = x_0^{({\rm G}_i)}\equiv  {1 - \tilde{c}^{({\rm G}_i)}\over 1-c} = {1\over c N^{({\rm G}_i)} + 1-c},
\label{eq:x0}
 \end{equation}
 where ${\rm G}_i$ is the group the species $i$ belongs to, either P or A. The superscript $(0)$ means the zeroth-order approximation. The first-order correction will be presented as well in the next subsection and then we move to the approximation for general $m$.
For $0<c<1$, Eq.~(\ref{eq:x0}) is positive for all $i$.  Therefore all the eigenvalues of the Jacobian in Eq.~(\ref{eq:lambda}) with $\cS^{(0)}=\emptyset$ are negative, and the fixed point with Eq.~(\ref{eq:x0}) for all $i$ is the stable fixed point. The increase of the competition strength $c$ or of the number of species $N^{({\rm G}_i)}$ leads to the decrease of the  abundance $x_i^{*(0)}$.  

\subsection{Weak mutualism}
Suppose that the mutualism strength $m$ is not zero but small. Expanding the inverse $\bbB^{-1}$ as 
\begin{equation}
\bbB^{-1}=\left(\bbB_0+m \bbA\right)^{-1}=\bbB_0^{-1} \sum_{n=0}^\infty \left(-m\bbA \bbB_0^{-1}\right)^n
\label{eq:BinvExpand}
\end{equation}
with $\bbB_0$ given in Eq.~(\ref{eq:B0}), and utilizing the relations like $\APA \JAA = \KP \JPA$ with $K_{pp^\prime} =k_p \delta_{pp^\prime}$ a block in the degree matrix  
\begin{equation}
\bbK\equiv\KP \oplus \KA,
\label{eq:Kmat}
\end{equation}
 one can evaluate the first order term in $m$ in Eq.~(\ref{eq:BinvExpand}) to obtain the first-order approximation 
  \begin{equation}
 x_i^{*(1)} \simeq x_0^{({\rm G}_i)}\left[1 + {m\over 1-c} {1-\tilde{c}^{({\rm \bar{G}}_i)} \over 1-\tilde{c}^{({\rm G}_i)}} \left(k_i - \tilde{c}^{({\rm G}_i)} \langle k\rangle^{({\rm G}_i)}\right)\right]
 \label{eq:x1}
 \end{equation}
 with the mean degree $\langle k\rangle^{\rm (G)} \equiv {L\over \myNG}$. This first-order approximation works if $||\bbA \bbB_0^{-1}||$ is sufficiently small. See Appendix~\ref{sec:firstorder} for more details. 
 
The formula in Eq.~(\ref{eq:x1})  allows us to understand the origin of the ambivalent effects of mutualism on the species abundance as observed in Fig.~\ref{fig:model} (d).   The two terms in the parentheses in Eq.~(\ref{eq:x1}) represent the  mutualistic benefit of a species $i$ (in group ${\rm G}_i$) from its $k_i$  mutualistic partners in group $\bar{\rm G}_i$, and the competition  with other species in the same group ${\rm G}_i$  that also benefit from mutualism, respectively. Their difference  may be positive or negative depending on degree $k_i$. It is the species  with $k_i>\tcG \langle k\rangle^{({\rm G}_i)}$ that finds abundance increasing  with increasing  $m$; the abundance of the species with $k_i<\tcG \langle k\rangle^{({\rm G}_i)}$ decreases with $m$, as its mutualistic benefit is overwhelmed by the competition with the species in the same group to the extent proportional to $m$ in the first-order approximation.

One caveat  is that Eq.~(\ref{eq:x1}) can be negative depending on parameters, which suggests that the fixed point with all species surviving is unstable and that some species will turn out to have zero abundance in the stable fixed point.  This will be explored in Sec.~\ref{sec:sephase}. 

\subsection{Annealed approximation for general $m$}
Each term for $n\geq 1$ in Eq.~(\ref{eq:BinvExpand}) represents the sum of the influences of other species on the abundance of a species built up over the pathways involving $n$ mutualistic pairs. To analytically track such higher-order contributions,  we consider  the {\it annealed} adjacency matrix $\tA_{pa} \equiv {k_p k_a \over L}$, meaning the probability to connect $p$ and $a$ in the  network ensemble for a given degree sequence~\cite{PhysRevE.80.051127}, and equivalently 
\begin{equation}
\bbtA = L^{-1} \bbK \bbJ^{(1)} \bbK,
\label{eq:Atilde}
\end{equation} 
where $\bbK$ is the degree matrix introduced in Eq.~(\ref{eq:Kmat}) and $\bbJ^{(1)}$ contains the matrices of $1$'s  at the off-diagonal blocks as $\bbJ^{(1)} \equiv  \JPA \oplus \JAP$ with $J_{pa}=J_{ap}=1$ for all $p$ and $a$. Then, after some algebra utilizing the properties of the matrices of $1$'s  as detailed in Appendix~\ref{sec:annealed}, we find each term $\bbB_0^{-1} (-m \bbtA \bbB_0^{-1})^n$  reduced to  
\begin{align}
\bbB_0^{-1} \left(-m\bbtA \bbB_0^{-1}\right)^n & =
\begin{cases}
-{\tm^n\over 1-c} \bbtK \bbtJ^{(1)} \bbtK & \ {\rm for} \ n=1,3,5,\ldots, \\
-{\tm^n\over 1-c} \bbtK \bbtJ^{(0)} \bbtK & \ {\rm for} \ n=2,4,6,\ldots,
\end{cases} 
\label{eq:Bn}
\end{align}
where we introduced $\bbtJ^{(1)} \equiv {\JPA \over \sqrt{\NP \NA} } \oplus {\JAP \over \sqrt{\NP \NA}}$, $\bbtK \equiv {{\KP \over \langle k\rangle^{\rm (P)}} - \tcP \IP \over \sqrt{\xi^{\rm (P)} - \tcP}} \oplus {{\KA \over \langle k\rangle^{\rm (A)}}-\tcA \IA\over \sqrt{\xi^{\rm (A)} - \tcA}}$, and the rescaled mutualism strength $\tm$
\begin{equation}
\tilde{m}\equiv\frac{m}{1-c} \sqrt{ \langle k\rangle^{(P)} \langle k\rangle^{(A)}(\xi^{(P)}-\tilde{c}^{(P)}) (\xi^{(A)}-\tilde{c}^{(A)})}
\label{eq:mtilde}
\end{equation} 
with the ratio of the first two moments of the mutualism degree
\begin{equation}
\xi^{\rm (G)} \equiv {\langle k^2\rangle^{\rm (G)} \over {\langle k\rangle^{\rm (G)}}^2}
\label{eq:xi}
\end{equation}
quantifying the heterogeneity of degree~\cite{GanYan2017}. 
The rescaled mutualism $\tm$ is the key parameter governing the species abundance, capturing the effects of network structural heterogeneity on the species abundance.

All the terms for $n\geq 1$ in Eq.~(\ref{eq:BinvExpand}) are proportional to either $\bbtJ^{(0)}$ or $\bbtJ^{(1)}$, with $\tm^n$ in the coefficient. Consequently,  the sum of the influences of  interspecific interactions over all possible pathways in Eq.~(\ref{eq:BinvExpand}) is reduced to two infinite geometric series,  manifesting the advantage of the annealed approximation. Then the inverse matrix is expressed in a closed form as
\begin{equation}
\bbtB^{-1} = \bbB_0^{-1}  - {1\over 1-c}{\tm \over 1-\tm^2} \bbtK \left(\tm \bbtJ^{(0)} + \bbtJ^{(1)}\right) \bbtK.
\label{eq:Binv}
\end{equation}
Substituting Eq.~(\ref{eq:Binv}) in Eq.~(\ref{eq:xi*}), we obtain the  fixed point 
\begin{align}
\tx^*_i =x_0^{({\rm G}_i)} \left(1+\bk_i  \tm {\tm + \eta^{({\rm G}_i \bar{\rm G}_i)} \over 1-\tm^2}\right)
\label{eq:xfpfull}
\end{align}
with the rescaled degree  
\begin{equation}
\bar{k}_i\equiv {{k_i\over \langle k\rangle^{({\rm G}_i)}} - \tilde{c}^{({\rm G}_i)} \over \xi^{({\rm G}_i)} - \tilde{c}^{({\rm G}_i)}},
\label{eq:kbar}
\end{equation}
and the asymmetry factor 
\begin{equation}
\eta^{\rm (PA)} \equiv {1-\tcA \over 1-\tcP} \sqrt{\langle k\rangle^{\rm (P)} (\xiP - \tcP) \over \langle k\rangle^{\rm (A)} (\xiA - \tcA)} = {1\over \eta^{\rm (AP)}}.
\label{eq:theta}
\end{equation}

The formula in Eq.~(\ref{eq:xfpfull}) is the main result of the present work, representing the abundance of individual species under heterogeneous mutualistic interactions and uniform intra-group competition. It is the  cornerstone of the results that follow in the next sections. The abundance is given by a non-linear function of the rescaled mutualism $\tm$ in Eq.~(\ref{eq:mtilde}), revealing how the higher-order contributions of interspecific interactions are combined with the network structure. The increase of mean connectivity $\langle k\rangle^{\rm (P,A)}$ or the increase of the degree heterogeneity $\xi^{\rm (P,A)}$ enhances the rescaled mutualism strength. As $\tm$ increases, $\tx_i^*$ may increase or decrease, depending on the sign of the rescaled degree $\bar{k}_i$. The rescaled degree quantifies the imbalance of the mutualism benefit  and the competition cost; $\tx_i^*$ increases (decreases) with $\tm$ if $\bk_i$ is positive (negative) as long as $0<\tm<1$. 

One can notice that $\tx_i^*$ in Eq.~(\ref{eq:xfpfull}) can be negative for some species $i$ depending on parameters and degree, implying then that Eq.~(\ref{eq:xfpfull}) is not the stable fixed point and invoking the necessity to classify correctly surviving and extinct species. The divergence of Eq.~(\ref{eq:xfpfull}) at $\tm=1$ suggests the onset of instability, which can be suppressed  up to a larger value of $\tm$  than one, along with the extinction of selected species as we will see.

\section{Phase diagram}

Some of the formulated abundances in Eq.~(\ref{eq:xfpfull}) can be negative depending on parameters. Then, by Eq.~(\ref{eq:lambda}), some species may have to have zero abundance and the remaining surviving species should have positive abundances different from Eq.~(\ref{eq:xfpfull}) as the interspecific interaction among the surviving species should be considered to formulate their abundances. In this section, we investigate the stable fixed point depending on parameters by identifying the species to go extinct, if any, and recalculating the abundance of the surviving species, and we obtain the phase diagram.

\subsection{Full coexistence phase} 
\label{sec:fcphase}

Let us call it  {\it full coexistence}  if there is no extinct species, i.e., if $x_i^{\rm (st)}>0$ for all $i$. For sufficiently small $\tm$, the numerically and analytically obtained values for the stationary-state abundance, $x_i^{\rm (st)}$ and $\tx_i^*$ in Eq.~(\ref{eq:xfpfull}), are in good agreement.
This agreement implies that  in the full-coexistence phase i)  the annealed approximation works, $x_i^{\rm (st)}\simeq  \tx_i^{\rm (st)}$, and  ii)  Eq.~(\ref{eq:xfpfull}) is stable, $ \tx_i^{\rm (st)}=\tx_i^*$,  where $\tx_i^{\rm (st)}$ is the stationary-state abundance from the solution to Eq.~(\ref{eq:LV}) with the annealed adjacency matrix $\tA_{ij}$ used.  See Appendix~\ref{sec:fcappendix} for different kinds of species abundances used in this paper. 

From Eq.~(\ref{eq:lambda}),  the full-coexistence fixed point in Eq.~(\ref{eq:xfpfull})  is stable only  if all $\tx_i^*$'s are positive. This holds for  $\tm<\tm_e^*(c) \equiv \min\left(\tm_e^{\rm *(P)}(c),\tm_e^{\rm *(A)}(c)\right)$ with 
 \begin{align}
&\tm_e^{\rm *(G)} (c) \equiv \nonumber\\
&\left\{
\begin{array}{ll}
1& \ {\rm for} \ c<c_{\rm min}^{\rm *(G)},\\
{\sqrt{4\left(1- \bk_{\rm min}^{\rm (G)}\right)  +  \left(\bk_{\rm min}^{\rm (G)} \eta^{\rm (G\bar{G})}\right)^2} +\bk_{\rm min}^{\rm (G)}  \eta^{\rm (G\bar{G})} \over 2\left(1- \bk_{\rm min}^{\rm (G)} \right)} & \ {\rm for} \ c>c_{\rm min}^{\rm *(G)}.
\end{array}
\right.
 \label{eq:tm*}
 \end{align}
Here  $\bar{k}_{\rm min}^{\rm (G)}$ is the rescaled degree of the group-G species having the smallest degree $k_{\rm min}^{\rm (G)}$, and $c_{\rm min}^{\rm *(G)}$  is defined as 
\begin{equation}
c_{\rm min}^{\rm *(G)} \equiv {k_{\rm min}^{\rm (G)}\over \myNG \left(\langle k\rangle^{\rm (G)} - k_{\rm min}^{\rm (G)}\right) + k_{\rm min}^{\rm (G)}},
\label{eq:c*}
\end{equation}
such that $\bk_{\rm min}^{\rm (G)}<0$ for $c>c_{\rm min}^{\rm (G)}$. 
 
\begin{figure}
\centering
\includegraphics[width=\columnwidth]{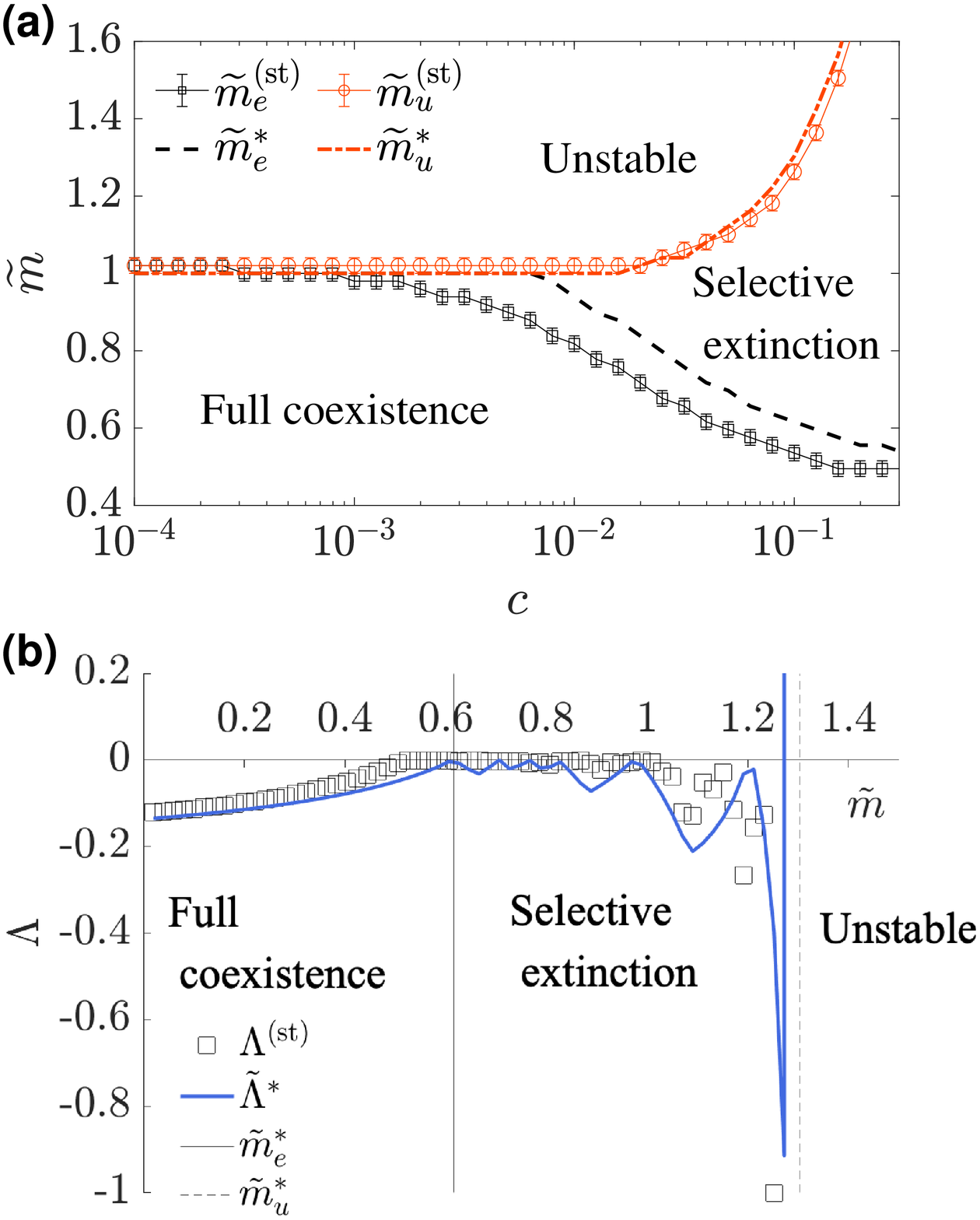}
\caption{Phase diagram and stability for the selected community~\cite{https://doi.org/10.1002/j.1537-2197.1982.tb13237.x}. (a)  The phase boundaries $\tm_e^{\rm (st)}$ and $\tm_u$ based on the stationary-state abundances are compared with the theoretical prediction $\tm_e^*$ and $\tm_u^*$ from  Eq.~(\ref{eq:tm*}) and the condition $\tm^{(+)}|_{\tm_u^*}=1$, respectively. (b) The largest real part $\Lambda^{\rm (st)}$ of the eigenvalues of the Jacobian matrix at $\vec{x}^{\rm (st)}$ and that at $\vec{x}^*$ approximated as $\tLambda^* = \max_{i\in \cS^{(0)}, j\in \cS^{(+)}} \left(\tx_i^{*\prime}, -\tx_j^*,-1\right)$ are shown as functions of $\tm$ for $c=0.1$.}
\label{fig:phase}
\end{figure}

The predicted boundary of the full coexistence phase $\tm_e^*(c)$ is shown by a dashed line in Fig.~\ref{fig:phase} (a), which is fixed at one for $c<c_{\rm min}^*\equiv \min(c_{\rm min}^{\rm *(P)}, c_{\rm min}^{\rm *(A)})$ and decreases with $c$ for $c>c_{\rm min}^*$. For $c< c_{\rm min}^*$, all species have positive rescaled degrees, $\bk_i>0$, and their fixed-point abundances $\tx_i^*$ increase with $\tm$ until diverging at $\tm=\tm_e^*(c)=1$.  Considering the fraction of diverging-abundance species $r_u\equiv S^{-1}\sum_i \theta(x_i^{\rm (st)}-M)$ with $M=10^5$, we call it the {\it unstable} phase if $r_u>0$. We define the instability threshold $\tm_u(c)$ such that $r_u>0$ for $\tm\geq \tm_u(c)$. One finds in Eq.~(\ref{eq:xfpfull}) that  the theoretical prediction for the instability threshold is given by $\tm_u^*(c)=1$ for $c<c_{\rm min}^*$, which agrees with the boundary of the full coexistence phase $\tm_e^*(c)=1$ in Eq.~(\ref{eq:tm*}). Notice that the unstable phase meets the full coexistence phase at $\tm=1$ for $c<c_{\rm min}^*$ [Fig.~\ref{fig:phase} (a)]. 

When $c$ is larger than $c_{\rm min}^*$ and $\tm$ is larger than $\tm_e^*(c)$, there appear some species $i$ with $\tx_i^*<0$ according to Eq.~(\ref{eq:xfpfull}), implying that they should go extinct, having zero abundance  in the true stable fixed point.  Computing   the fraction of extinct species in terms of their stationary-state abundances as 
\begin{equation}
r_e^{\rm (st)}\equiv S^{-1} \sum_{i=1}^S \theta(\epsilon - x_i^{\rm (st)}),
\label{eq:rest}
\end{equation}
with $\epsilon = 10^{-5}$ introduced under finite precision of numerics and the Heaviside step function $\theta(x)=1$ for $x> 0$ and $0$ otherwise, we find that $r_e^{\rm (st)}$ becomes non-zero as $\tm$ exceeds the extinction threshold $\tm_e^{\rm (st)}(c)$ for $c>c_{\rm min}^*$, which is well approximated by the predicted threshold $\tm_e^*(c)$ in Eq.~(\ref{eq:tm*}) [Fig.~\ref{fig:phase} (a)]. Let us call it {\it selective extinction}  if there exist extinct species ($r_e^{\rm (st)}>0$) but no abundance-diverging species ($r_u=0$). Our analysis shows that the full coexistence phase meets the selective extinction phase at $\tm_e^{\rm (st)}\simeq \tm_e^*$ for $c>c_{\rm min}^*$. While we showed that the full-coexistence phase ends at $\tm_e^{\rm (st)}$, it remains to be explored which species go extinct and what happens for the remaining surviving species. It will be addressed in the next subsection in details. 
 
In Fig.~\ref{fig:phase} (b), the largest real part $\Lambda^{\rm (st)}$ of the eigenvalues of the Jacobian $\bbH$ computed at the stationary-state abundance  $\vec{x}^{\rm (st)}=(x^{\rm (st)}_i)$ is shown to be negative in the full coexistence phase, demonstrating the stability of  $\vec{x}^{\rm (st)}$. The analytically-obtained fixed point $\vec{\tx}^*=(\tx_i^*)$  is stable as well; The largest eigenvalue $\tLambda^*$ of the Jacobian $\tilde{H}_{ij}= \delta_{ij} \left[1 + \sum_{\ell=1}^S \tilde{B}_{i\ell} \tx_\ell^*\right] + \tx_i^* \tilde{B}_{ij}$  is evaluated as $\tLambda^*= \max_i (-\tx_i^*,-1)$, from Eq.~(\ref{eq:lambda}) and the property that $\tilde{H}_{ij} $ has the eigenvalue $-1$ [Appendix~\ref{sec:eigen}], and remains negative in the full-coexistence phase. Moreover,   we see a good agreement between $\Lambda$ and $\tLambda^*$.

\subsection{Selective extinction phase} 
\label{sec:sephase}

If the set $\cS^{(0)}$ of extinct species  is known,  one can apply Eq.~(\ref{eq:xfpfull}) to the subcommunity of the surviving species and obtain their abundances analytically. Removing the rows and columns of the species belonging to $\cS^{(0)}$  in the full matrix $\bbtB$ and also in the adjacency matrix $\bbtA$, one can obtain the effective interaction matrix $\bbtB^{(+)}$ and the effective adjacency matrix $\bbtA^{(+)}$ for the surviving-species community, from which we can compute the effective quantities such as $x_0^{\rm (G,+)}$, $\bk_i^{(+)}$, $\tm^{(+)}$, and $\eta^{\rm (PA,+)}$ as described in Appendix~\ref{sec:effective}.  Using them in Eq.~(\ref{eq:xfpfull}), one can obtain the approximate stable fixed point 
\begin{equation}
\tx^*_i = 
\begin{cases}
0 & \ {\rm for} \ i\in \cS^{(0)}, \\
x_0^{({\rm G}_i,+)} \left(1+\bk_i^{(+)}  \tm^{(+)} {\tm^{(+)} + \eta^{({\rm G}_i \bar{\rm G}_i,+)} \over 1-(\tm^{(+)})^2}\right) &  \ {\rm for} \ i\in \cS^{(+)}.
\end{cases}
\label{eq:xfpfulleff}
\end{equation}
The sets of extinct and surviving species, $\cS^{(0)}_{\rm stable}$ and $\cS^{(+)}_{\rm stable}$  for the {\it stable} fixed point are  not given a priori. Examining all possible sets $\cS^{(0)}$ and identifying the one yielding all negative eigenvalues as given in Eq.~(\ref{eq:lambda}) could be done but  takes a very long time. 

Our analytic results, Eqs.~(\ref{eq:lambda}) and (\ref{eq:xfpfulleff}), give a clue to proceed. Suppose that we have a pair of disjoint sets $\cS^{(0)}$ and $\cS^{(+)}$ with $\cS\equiv \cS^{(0)}\cup \cS^{(+)}$ the set of all species.  If a species in $\cS^{(+)}$ with a small (large) effective degree has a negative (positive) value of $\tx_i^*$ from Eq.~(\ref{eq:xfpfulleff}), then it will be likely to be in the right set $\cS^{(0)}_{\rm stable}$ ($\cS^{(+)}_{\rm stable}$) for the stable fixed point according to Eq.~(\ref{eq:lambda}).  Using this reasoning, we can update iteratively $\cS^{(0)}$ and $\cS^{(+)}$ towards obtaining $\cS^{(0)}_{\rm stable}$ and $\cS^{(+)}_{\rm stable}$ as follows. \\

Initially we begin with $\cS^{(0)}=\emptyset$, $\cS^{(+)}=\cS$, $\bbtB^{(+)}=\bbtB$, and $\tx_i^*$ evaluated by Eq.~(\ref{eq:xfpfulleff}). Then, the following procedures are repeated until stopping at the step (iii): \\
(i) We label as new extinct species all the plant (animal) species $p_{e}$'s ($a_{e}$'s) with their fixed-point abundances $\tx_{p_{e}}^{\rm *}$ ($\tx_{a_{e}}^{\rm *}$) having {\it different} sign from that  of the hub plant species $\tx_{p_{\rm hub}}^*$ (animal species $\tx_{a_{\rm hub}}^*$), the one having the largest effective degree. \\
(ii) Go to step iv) if there are such new extinct species, or \\
(iii) Stop if there are none.\\
(iv) We remove those new extinct species from $\cS^{(+)}$ and add them to $\cS^{(0)}$ and update $\bbtB^{(+)}$ by eliminating their rows and columns, and\\
(v) Set $\tx_{p_{e}}^{*}=\tx_{a_{e}}^{*}=0$, and evaluate $\tx_i^*$'s for the remaining surviving species $i$ by Eq.~(\ref{eq:xfpfulleff}) with using the new $\bbtB^{(+)}$.

\begin{figure*}
\centering
\includegraphics[width=2\columnwidth]{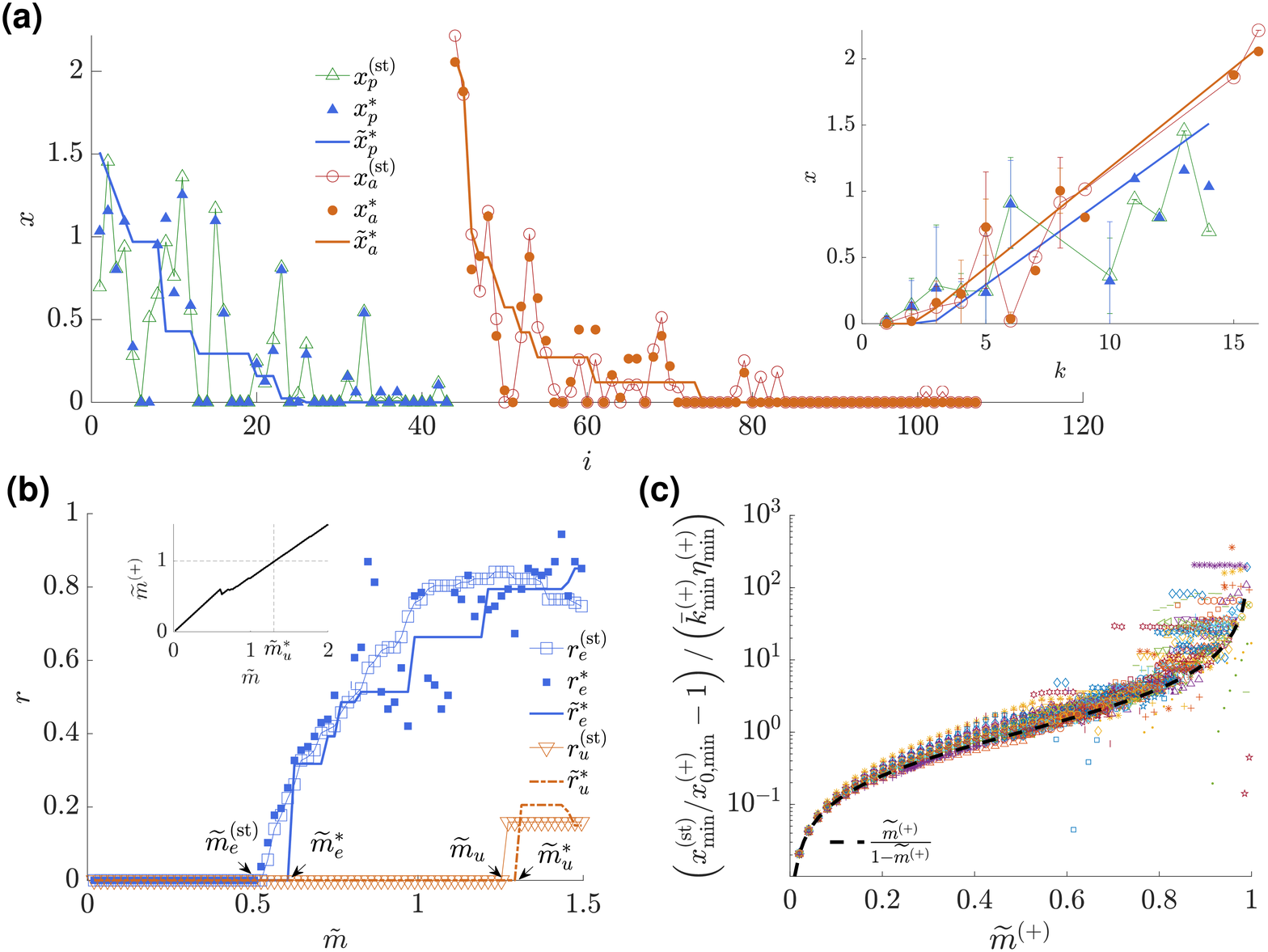}
\caption{Selective extinction phase.
(a)  Stationary-state abundances $x_i^{\rm (st)}$ are compared with the stable fixed-point ones $x_i^*$'s and $\tx_i^*$'s for $c=0.1$ and $\tm=0.8$. The species index $i$ is arranged in the descending order of degree among plants and among animals. Inset: Abundance vs. degree. 
(b) Fraction $r_e^{\rm (st)}$ of extinct species  and  $r_u$ of the abundance-diverging species based on the stationary-state abundance for $c=0.1$. They are compared with $\tr_e^*$ and  $\tr_u^*$ based on the stable fixed point $\tx_i^*$'s and also with $r_e^*$ based on $x_i^*$'s.   The critical points are also marked, $\tm_e^{\rm (st)}\simeq 0.53$ and $\tm_e^*\simeq 0.61$. Inset: The effective rescaled mutualism strength $\tm^{(+)}$ reaches $1$ at $\tm = \tm_u^*\simeq 1.3$ for $c=0.1$.  
(c) The collapse of the non-zero minimum abundances $x_{\rm min}^{\rm (st)}$ rescaled as  in Eq.~(\ref{eq:scal}) in 46 real-world communities as functions of  $\tm^{(+)}$. }
\label{fig:extinction}
\end{figure*}

Note that the effective rescaled mutualism strength $\tm^{\rm (+)}$ can be larger than  one, making the hub abundances negative according to Eq.~(\ref{eq:xfpfulleff}) in the middle of  the above procedures, which is why we compare the sign of the abundance of a species with that of hub species to identify extinct species.   At the end of these procedures we are given $\cS^{(0)}_{\rm stable}$ and $\cS^{(+)}_{\rm stable}$, which we use to obtain the  stable fixed point  $\tx_i^*$'s from Eq.~(\ref{eq:xfpfulleff})~\cite{githubcode}.

As shown in Fig.~\ref{fig:extinction} (a), the predicted abundance $\tx_i^*$ approximates reasonably the stationary-state abundance $x_i^{\rm (st)}$. It grows with degree $k_i$ and takes a zero value for the species with the smallest degrees, demonstrating the crucial role of degree on extinction. Its origin can be understood by examining the rescaled degree, the imbalance of the mutualistic benefit and the competition cost, the former of which is proportional to the raw degree. The stable fixed point  $\tx_i^*$ predicts  whether a species survives or goes extinct correctly for 80\% of species across parameters; The predicted fraction $\tilde{r}_e^* \equiv S^{-1}\sum_i \theta(\epsilon - \tx_i^*)$  of extinct species is in good agreement with the true value $r_e^{\rm (st)}$ in Eq.~(\ref{eq:rest}), which plays  the role of the order parameter distinguishing the full-coexistence phase ($r_e=0$) and the selective extinction phase ($r_e>0$) [Fig.~\ref{fig:extinction} (b)]. 

Deviations stem from the annealed approximation; The stationary-state abundance $\tx_i^{\rm (st)}$ under the annealed adjacency matrix agrees perfectly with $\tx_i^*$ (Appendix~\ref{sec:abundanceannealed}), and the predicted fraction of extinct species $\tilde{r}_e^*$ explains very well the true value $\tilde{r}_e$ under the annealed adjacency matrix  (Appendix~\ref{sec:accuracy}).  Instead of using $\bbtB$ and Eq.~(\ref{eq:xfpfulleff}), one can also use $\bbB$ and Eq.~(\ref{eq:xi*}) in the above procedures to identify the set of extinct and surviving species and obtain the stable fixed point $x_i^*$ from Eq.~(\ref{eq:xi*}), which agrees very well with $x_i^{\rm (st)}$ as shown in Fig.~\ref{fig:extinction} (a). 

One might have expected instability to arise at $\tm=1$ from Eq.~(\ref{eq:xfpfull}). However the extinction of the small-degree species effectively reduces $\tm$ to $\tm^{\rm (+)}$ [Fig.~\ref{fig:extinction} (b)] and the abundances of the surviving species are evaluated by Eq.~(\ref{eq:xfpfulleff}). The effective rescaled mutualism strength $\tm^{(+)}$ is kept smaller than $1$, preventing the onset of instability, up to $\tm_u(c)$  for $c>c_{\rm min}^*$. As the smallest-degree species go extinct,  we find that the degree heterogeneity is reduced in the interaction network of the surviving species, which drives the reduction of the  effective rescaled mutualism strength (Appendix~\ref{sec:effective}). Like in  the full coexistence phase,  the largest real part of the eigenvalues $\Lambda$  at $\vec{x}^{\rm (st)}$ and $\tLambda^* = \max_{i\in \cS^{(0)}_{\rm stable}, j\in \cS^{(+)}_{\rm stable}} \left(\tx_i^{*\prime}, -\tx_j^*,-1\right)$ at the stable fixed point $\tx_i^*$ remain negative in the selective extinction phase, demonstrating stability [Fig.~\ref{fig:phase} (b)].   

The selective extinction phase meets the unstable phase at $\tm_u(c)$ where the fraction of abundance-diverging species $r_u$ becomes non-zero. The instability threshold $\tm_u(c)$ coincides with its theoretical prediction $\tm_u^*$ at which $\tm^{(+)}=1$ and $\tx_i^*$ in   Eq.~(\ref{eq:xfpfulleff}) diverges [Fig.~\ref{fig:extinction} (b)]. For $\tm>\tm_u^*$, the non-zero components of the stable fixed point becomes negative, featuring the non-zero fraction $\tr_u^*\equiv S^{-1}\sum_i \theta(-\epsilon -\tx_i^*)$ of the species having negative $\tx_i^*$.

Lastly, to demonstrate the general applicability of our analytic results, we study the minimum stationary-state abundance of the surviving species $x_{\rm min}^{\rm (st)}$ in 46 large empirical mutualistic networks with $\NP, \NA\geq 20$~\cite{web_of_life}. As in Sec.~\ref{sec:model}, we assume the uniform intra-group competition with strength $c$, and we use the  data-sets to construct the mutualism adjacency matrices with the mutualism strength $m$.   

From the  theoretical framework developed in the previous sections,  $x_{\rm min}^{\rm (st)}$ can be approximated by the non-zero minimum component of the stable fixed point,  $\tx_{\rm min}^*=\min_{i\in \cS^{(+)}_{\rm stable}} \tx_i^*=\tx_{i_{\rm min}}^*$ representing the predicted abundance of species $i_{\rm min}$.  From Eq.~(\ref{eq:xfpfulleff}),  we find that $\tx_{\rm min}^*$  behaves as a function of $\tm^{(+)}$ as
\begin{equation}
{ {\tx_{\rm min}^{*}  \over x_{0, {\rm min}}^{(+)}}-1 \over \bk_{\rm min}^{(+)} \eta_{\rm min}^{(+)}}\simeq
 {\tm^{(+)}  \over 1-\tm^{(+)}},
 \label{eq:scal}
\end{equation}
where $\bk_{\rm min}^{(+)} \equiv\bar{k}_{i_{\rm min}}$, $x_{0,{\rm min}}^{(+)}\equiv x_0^{({\rm G}_{i_{\rm min}},+)}$, and $\eta_{\rm min}^{(+)}\equiv\eta^{({\rm G}_{i_{\rm min}} \bar{\rm G}_{i_{\rm min}},+)}$, and we approximate $1+ \tm^{(+)}/\eta^{(+)}_{\rm min}$ by $1+\tm^{(+)}$ in the right-hand-side.  In Fig.~\ref{fig:extinction} (c), the plots of the rescaled minimum abundance given by the left-hand-side of Eq.~(\ref{eq:scal}) with $x_{\rm min}^{\rm (st)}$ in place of $\tx_{\rm min}^*$ versus $\tm^{(+)}$ for $c=0.1$ in 46 empirical mutualistic networks collapse reasonably onto ${\tm^{\rm (+)} \over 1-\tm^{(+)}}$ in agreement with Eq.~(\ref{eq:scal}). There are outliers, though. About 6\% of the data points have their rescaled minimum abundances negative and they are thus neglected in Fig.~\ref{fig:extinction} (c). Some of the outliers seen in Fig.~\ref{fig:extinction} (c) are attributed to the annealed approximation, which return close to the theoretical curve in the annealed network [see Fig.~\ref{fig:Eff_annealed} (e)]. Outliers seen for $\tm^{(+)}\simeq 1$ in Fig.~\ref{fig:Eff_annealed} (e) suggest the possible relevance of network characteristics beyond the degree sequence, which needs further investigation.
 
\section{Summary and discussion} 
While various theoretical approaches have been established for studying the stability and biodiversity of random unstructured communities,  the relation between the structured  interaction, universal in the real world, and  the emergent phenomena of the community has been little understood, partly due to the lack of an analytically tractable model. Here we considered a model community of two groups of species - plants and pollinators -  under uniform intra-group competition and empirical heterogeneous inter-group mutualism, and we investigated analytically and numerically the abundance and extinction of individual species in that community. 

Deriving the stability condition and the stable fixed point of the LV equation, we quantified the influences of the structural heterogeneity.  The strength of mutualism is rescaled by the degree heterogeneity.  The species with  few mutualistic partners are driven to extinction by their little benefit of mutualism compared with the high cost of competition.  As the mutualism strength increases, more species find benefit falling short of cost, resulting in the increase of the number of extinct species.   The effective rescaled mutualism among the surviving species is reduced with respect to  the original one, which enables the community of the surviving species to be stable for a wide range of parameters, delaying the onset of instability. The number of extinct species and the fraction of the abundance-diverging species play the roles of the order parameter in the phase diagram.

Going beyond the annealed approximation to identify further network characteristics  than degree may provide rich concepts and methods to characterize the structure-function relationship of ecological communities. Contrary to the unstructured communities where the interspecific interaction patterns and strengths are random and distributed identically across species, the number of mutualistic partners  is different from species to species  in the structured community that we study in the present work. We showed how the abundance and the likelihood of extinction depend on the  degree of a species. Real-world communities should exhibit both non-uniform interaction strengths and heterogeneous connection patterns. If our analytic framework can be generalized to handle not only heterogeneity but also the randomness of the interaction matrix,  it will help us to better understand how structural heterogeneity and randomness together govern the stability and species extinction of real-world communities.

\section*{Acknowledgment}
We thank Matthieu Barbier, Hye Jin Park, Yongjoo Baek, and Sang Hoon Lee for valuable comments.  This work was supported by the National Research Foundation of Korea (NRF) grants funded by the Korean Government (No. 2019R1A2C1003486 (D.-S.L) and No. 2020R1A2C1005334 (J.W.L)) and a KIAS Individual Grant (No. CG079901) at Korea Institute for Advanced Study (D.-S.L).

\appendix 

\section{Properties of the matrices of $1$'s }
\label{sec:Jmat}

The matrix of $1$'s  denoted by ${\bf J}$ having all elements equal to $1$, $J_{ij}=1$ for all $i$ and $j$, is  used in the present work to represent the all-to-all uniform competition among plants  and among animals  via $\JPP$ and  $\JAA$ of dimension $\NP\times \NP$ and $\NA \times \NA$, respectively, and also to represent the uniform coupling between plants and animals, appearing in the annealed adjacency matrix, via $\JPA$ and $\JAP$ of dimension $\NP\times \NA$ and $\NA \times \NP$, respectively. We also consider its integrated versions in block-matrix form
\begin{align}
\bbJ^{(0)} &\equiv 
\begin{pmatrix}
\JPP & {\bf 0} \\
{\bf 0} & \JAA
\end{pmatrix}
= \JPP \oplus \JAA, \nonumber\\
\bbJ^{(1)} &\equiv 
\begin{pmatrix}
{\bf 0} & \JPA \\
\JAP & {\bf 0}
\end{pmatrix}
= \JPA \oplus \JAP,
\end{align}
with $\oplus$ representing the direct sums of two matrices of $1$'s  defined on different groups of nodes, and
the rescaled matrices given by
 \begin{align}
\bbtJ^{(0)} &\equiv 
\begin{pmatrix}
{\JPP \over \NP} & {\bf 0} \\
{\bf 0} & {\JAA \over \NA}
\end{pmatrix}
= {\JPP\over \NP} \oplus {\JAA\over \NA}, \nonumber\\
\bbtJ^{(1)} &\equiv 
\begin{pmatrix}
{\bf 0} & {\JPA\over \sqrt{\NP\NA}} \\
{\JAP\over \sqrt{\NP\NA}} & {\bf 0}
\end{pmatrix}
= {\JPA \oplus \JAP\over \sqrt{\NP\NA}}.
\label{eq:scaledJ}
\end{align}
In this appendix, we present their useful properties, which are used to derive the analytic results presented in the main text.

Let us denote the matrices of $1$'s  of dimension $N_1 \times N_2$ by ${\bf J}^{(N_1\times N_2)}$.  If one multiplies ${\bf J}^{(N_1\times N_2)}$ and ${\bf J}^{(N_2\times N_3)}$, then  she obtains 
${\bf J}^{(N_1\times N_2)} {\bf J}^{(N_2\times N_3)} = N_2 {\bf J}^{\rm (N_1 \times N_3)},$ since $\sum_{j=1}^{N_2} J_{ij}^{\rm (N_1\times N_2)} J_{jk}^{\rm (N_2\times N_3)} = \sum_{j=1}^{N_2} 1 = N_2$ for all $1\leq i\leq N_1$ and $1\leq k\leq N_3$. Therefore we have 
\begin{equation}
{\bf J}^{\rm (G_1 G_2)} {\bf J}^{\rm (G_2 G_3)} = N^{\rm (G_2)} {\bf J}^{\rm (G_1 G_3)},
\end{equation}
where ${\rm G_1, G_2, G_3 \in \{P, A\}}$. 
Using this result, one can see that the rescaled block matrices of $1$'s  satisfy
\begin{align}
\bbtJ^{(0)} \bbtJ^{(0)} &= \bbtJ^{\rm (0)}, \nonumber\\
\bbtJ^{(1)} \bbtJ^{(1)} &= \bbtJ^{\rm (0)}, \nonumber\\
\bbtJ^{(0)} \bbtJ^{(1)} &= \bbtJ^{(1)} \bbtJ^{(0)} = \bbtJ^{\rm (1)}.
\end{align}

The multiplication of  ${\bf J}$  with the adjacency matrix $\APA$ or $\AAP ={\APA}^\intercal $ is evaluated as 
\begin{align}
&{\bf A}^{\rm (G_1 G_2)} {\bf J}^{\rm (G_2 G_3)} = {\bf K}^{\rm (G_1)} {\bf J}^{\rm (G_1 G_3)}, \nonumber \\
&{\bf J}^{\rm (G_1 G_2)} {\bf A}^{\rm (G_2 G_3)} = {\bf J}^{\rm (G_1 G_3)} {\bf K}^{\rm (G_3)},
\end{align}
where we use for instance that $\sum_{a^\prime} A_{pa^\prime} J_{a^\prime a} = \sum_{a^\prime} A_{pa^\prime}= k_p$ and $\sum_{p^\prime} J_{pp^\prime} A_{p^\prime a} = k_a$. Note that $K_{pp^\prime} = k_p \delta_{pp^\prime}$ and $K_{aa^\prime} = k_a \delta_{aa^\prime}$. The block adjacency matrix defined as
\begin{equation}
\bbA \equiv 
\begin{pmatrix}
{\bf 0} & \APA \\
\AAP& {\bf 0}
\end{pmatrix}
=\APA\oplus \AAP,
\end{equation}
 the block rescaled matrices of $1$'s  $\bbtJ^{(0)}$ and $\bbtJ^{(1)}$, and the block degree matrix defined as
 \begin{equation}
\bbK \equiv 
\begin{pmatrix}
\KP & {\bf 0}  \\
 {\bf 0} & \KA
\end{pmatrix}
=\KP \oplus \KA
\end{equation}
 satisfy the following equalities
\begin{align}
\bbA \bbtJ^{(0)} &= \bbK\bbtN \bbtJ^{(1)}, \ \bbtJ^{(0)} \bbA =  \bbtJ^{(1)}\bbtN \bbK,  \nonumber\\
\bbA \bbtJ^{(1)} &=   \bbK \bbtN \bbtJ^{(0)}, \ \bbtJ^{(1)} \bbA = \bbtJ^{(0)}\bbtN \bbK 
\label{eq:AJ}
\end{align} 
with $\bbtN \equiv \begin{pmatrix}\sqrt{\NP\over \NA}\IP & {\bf 0} \\ {\bf 0} & \sqrt{\NA\over \NP} \IA\end{pmatrix} = \sqrt{\NP\over \NA}\IP \oplus \sqrt{\NA\over \NP} \IA$.

Multiplying the degree matrices $\KP$ and $\KA$ by ${\bf J}$ matrices yields
\begin{align}
&{\bf J}^{\rm (G_1G_2)} {\bf K}^{\rm (G_2)} {\bf J}^{\rm (G_2G_3)} = L {\bf J}^{\rm (G_1 G_3)}, \nonumber \\
&{\bf J}^{\rm (G_1G_2)}\left({\bf K}^{\rm (G_2)}\right)^2 {\bf J}^{\rm (G_2G_3)} = N^{\rm (G_2)} \langle k^2\rangle^{\rm (G_2)} {\bf J}^{\rm (G_1 G_3)},
\end{align}
where we used $\sum_{a_1 a_2} J_{a a_1} K_{a_1 a_2} J_{a_2 p} = \sum_{a_1} k_{a_1} = L$ and $\sum_{p_1, p_2, p_3} J_{pp_1} K_{p_1 p_2} K_{p_2 p_3} J_{p_3 a} = \sum_{p_1} k_{p_1}^2 = \NP \langle k^2\rangle^{\rm (P)}$, and $\langle k^2\rangle^{\rm (G)} = \sum_{i\in \cS^{\rm (G)}} k_i^2 /N^{\rm (G)}$ is the mean of the square of the degree of species of group G with $\cS^{\rm (G)}$ the set of group-G species. The block matrices satisfy
\begin{align}
\bbtJ^{(0)} \bbK \bbtJ^{(0)} &= \bbtJ^{(0)} \langle \bbK \rangle,\nonumber\\
\bbtJ^{(0)} \bbK \bbtJ^{(1)} &= \langle \bbK \rangle\bbtJ^{(1)},  \nonumber\\
\bbtJ^{(1)} \bbK \bbtJ^{(0)} &= \bbtJ^{(1)} \langle \bbK \rangle, \nonumber\\
\bbtJ^{(1)} \bbK \bbtJ^{(1)} &= \langle k\rangle^{\rm (P)} \langle k\rangle^{\rm (A)} \langle \bbK \rangle^{-1} \bbtJ^{(0)},\nonumber\\
\bbtJ^{(1)} \bbK^2 \bbtJ^{(1)} &= \langle k^2\rangle^{\rm (P)} \langle k^2\rangle^{\rm (A)} \langle \bbK^2 \rangle^{-1} \bbtJ^{(0)},
\label{eq:KJ}
\end{align}
with $\langle \bbK \rangle \equiv \langle k\rangle^{\rm (P)} \IP + \langle k\rangle^{\rm (A)}\IA$ being the sum of the identity matrices multiplied by the group averages. These relations are valid also for a function $f(\bbK)$ of $\bbK$ as 
\begin{align}
\bbtJ^{(0)} f(\bbK) \bbtJ^{(0)} &= \bbtJ^{(0)} \langle f(\bbK) \rangle,\nonumber\\
\bbtJ^{(0)} f(\bbK) \bbtJ^{(1)} &= \langle f(\bbK) \rangle\bbtJ^{(1)}, \nonumber\\
\bbtJ^{(1)} f(\bbK) \bbtJ^{(0)} &= \bbtJ^{(1)} \langle f(\bbK) \rangle, \nonumber\\
\bbtJ^{(1)} f(\bbK) \bbtJ^{(1)} &= \langle f(\bbK)\rangle^{\rm (P)} \langle f(\bbK)\rangle^{\rm (A)} \langle f(\bbK) \rangle^{-1} \bbtJ^{(0)},
\label{eq:fJ}
\end{align}
where 
\begin{align}
\langle f(\bbK)\rangle &\equiv {{\rm Tr} \,f(\KP) \over \NP} \IP \oplus  {{\rm Tr}\,  f(\KA) \over \NA} \IA \nonumber\\
&= \langle f(\bbK)\rangle^{\rm (P)} \IP \oplus\langle f(\bbK)\rangle^{\rm (A)} \IA,
\label{eq:fullave}
\end{align}
is the sum of the group averages of $f(\bbK)$, and its inverse means $\langle f(\bbK)\rangle^{-1} = {1\over \langle f(\bbK)\rangle^{\rm (P)}} \IP \oplus {1\over \langle f(\bbK)\rangle^{\rm (A)} }\IA$. For general $z$,  $\langle f(\bbK)\rangle^z = \left(\langle f\rangle^{\rm (P)}\right)^z \IP \oplus \left(\langle f\rangle^{\rm (A)}\right)^z \IA$. 
The multiplication of $\langle f(\bbK)\rangle$ and $\bbtJ^{(1)}$ is not commutative:
\begin{align}
\langle f(\bbK)\rangle \bbtJ^{(1)} &= \langle f(\bbK)\rangle^{\rm (P)}  \langle f(\bbK)\rangle^{\rm (A)} \bbtJ^{(1)} \langle f(\bbK)\rangle^{-1},  \nonumber\\
 \bbtJ^{(1)} \langle f(\bbK)\rangle&= \langle f(\bbK)\rangle^{\rm (P)}  \langle f(\bbK)\rangle^{\rm (A)} \langle f(\bbK)\rangle^{-1} \bbtJ^{(1)},
\end{align} 
which can be seen by considering for instance the P-block of $\langle f(\bbK)\rangle \bbtJ^{(1)}\mathbb{X}$ as $\langle f\rangle^{\rm (P)} \JPA {\bf X}^{\rm (A)} = \langle f\rangle^{\rm (P)} \langle f\rangle^{\rm (A)} \JPA {1\over \langle f\rangle^{\rm (A)}} {\bf X}^{\rm (A)}$ with $\mathbb{X} = {\bf X}^{\rm (P)} \oplus {\bf X}^{\rm (A)}$.

\section{Eigenvalues of the Jacobian matrix}
\label{sec:eigen}

To help understand Eq.~(\ref{eq:lambda}), here we construct the Jacobian matrix for a small community as a concrete example and present the eigenvalue perturbation theory that we used in the main text.

\subsection{Jacobian matrix for a small community} 
Let us consider a community consisting of $S=4$ species and a fixed point $\vec{x}^* = (x_1^*\neq 0 , 0, 0, x_4^*\neq 0)$ of the LV equation for the community, which corresponds to $\cS^{(0)}=\{2,3\}$ and $\cS^{(+)} = \{1,4\}$.  As shown in Eq.~(\ref{eq:xi*}), the non-zero components (abundances) of species $1$ and $4$ in $\cS^{(+)}$ satisfy 
\[\bbB^{(+)} \begin{pmatrix}x_1^* \\ x_4^*\end{pmatrix} = \begin{pmatrix} B_{11} & B_{14}\\ B_{41} & B_{44}\end{pmatrix} \begin{pmatrix}x_1^* \\ x_4^*\end{pmatrix}= - \begin{pmatrix} 1\\ 1\end{pmatrix},\] where we introduced  the effective interaction matrix $\bbB^{(+)}$ by eliminating the rows and columns of the species, $2$ and $3$, of $\cS^{(0)}$ in the original $4\times 4$ interaction matrix $\bbB$ while keeping the original indices of rows and columns. Using Eq.~(\ref{eq:Hij}) and rearranging the order of indices as $(2\ 3 \  1\  4)$, we find the Jacobian matrix at the considered fixed point given by 
\begin{align}
& \bbH = \nonumber\\
&\begin{pmatrix}
1 + B_{21} x_1^* + B_{24} x_4^* & 0 & 0 & 0 \\
0 & 1 + B_{31} x_1^* + B_{34} x_4^* & 0 & 0\\
x_1^* B_{12} & x_1^* B_{13} & x_1^*B_{11} & x_1^* B_{14}\\
x_4^* B_{42} & x_4^* B_{43} & x_4^*B_{41} & x_4^* B_{44}
\end{pmatrix},\nonumber
\end{align}
 which is represented as in Eq.~(\ref{eq:Jacobian}) in terms of the block matrices 
\begin{align}
{\bf H}^{(00)} &=\begin{pmatrix} H_{22} & 0 \\ 0 & H_{33}\end{pmatrix}\nonumber\\
&=\begin{pmatrix}1 + B_{21} x_1^* + B_{24} x_4^* & 0\\ 0 & 1 + B_{31} x_1^* + B_{34} x_4^*\end{pmatrix} \nonumber
\end{align}
and  
\[
{\bf H}^{(++)} = \begin{pmatrix} H_{11} & H_{14} \\ H_{41} & H_{44}\end{pmatrix} = \begin{pmatrix} x_1^*B_{11} & x_1^* B_{14}\\  x_4^*B_{41} & x_4^* B_{44}\end{pmatrix}.
\]
 Note that $B^{(+)}_{ij} = B_{ij}$ if $i,j$ = $1$ or $4$ in our example. 
 
The diagonal elements $H_{22}$ and $H_{33}$ are the eigenvalues of ${\bf H}^{(00)}$, as ${\bf H}^{(00)}$ is a diagonal matrix. The diagonal elements can be further simplified. Let us first consider  $H_{22}=1 + B_{21} x_1^* + B_{24} x_4^*$.  
A clue  is obtained  by considering a neighboring fixed point $\vec{x}^{*\prime}= (x_1^{*\prime}, x_2^{*\prime}, 0, x_4^{*\prime})$ with $\cS^{(0)\prime}=\{3\}$ and $\cS^{(+)\prime} = \{1,2,4\}$, where species $2$ has a non-zero component, as well as species $1$ and $4$. Their abundances satisfy 
\[
\begin{pmatrix} B_{11} & B_{12} & B_{14}\\ B_{21} & B_{22} & B_{24} \\ B_{41} & B_{42} & B_{44}\end{pmatrix} \begin{pmatrix}x_1^{*\prime} \\ x_2^{*\prime}\\ x_4^{*\prime} \end{pmatrix}= -\begin{pmatrix} 1\\ 1 \\ 1\end{pmatrix}.
\] 
Among the three equalities from this equation is   $B_{21} x_1^{*\prime} + B_{22} x_2^{*\prime} + B_{24} x_4^{*\prime} = -1$. Rearranging the terms and recalling $B_{22}=-1$, we find that $x_2^{*\prime} = 1 + B_{21} x_1^{*\prime} + B_{24} x_4^{*\prime}$. If we assume that $x_1^{*\prime} \simeq x_1^*$ and $x_4^{*\prime} \simeq x_4^*$, i.e., that the non-zero components of species $1$ and $4$ are similar between the two fixed points $\vec{x}^*$ and $\vec{x}^{*\prime}$, then we find that \[H_{22} = 1 + B_{21} x_1^* + B_{24} x_4^*  \simeq 1 + B_{21} x_1^{*\prime} + B_{24} x_4^{*\prime} = x_2^{*\prime}.\] The assumption is expected to be valid if $S$ is large and the number of species having non-zero components is sufficiently large at the fixed point, for which allowing one more species to have a non-zero component would not change much the non-zero components of other species.  Similarly, one can simplify $H_{33}$ as $H_{33} \simeq x_3^{*\prime}$ by using another neighboring fixed point corresponding to $\cS^{(0)\prime} = \{2\}$ and $\cS^{(+)\prime} = \{1, 3,4\}$.  In general, one can obtain the approximate expression $H_{ii} \simeq x_i^{*\prime}$ for every $i\in \cS^{(0)}$ by considering the  neighboring fixed point corresponding to $\cS^{(0)\prime} = \cS^{(0)} - \{i\}$ and $\cS^{(+)\prime} = \cS^{(+)}\cup \{i\}$. Therefore, given a fixed point corresponding to $\cS^{(0)}$ and $\cS^{(+)}$, $x_i^{*\prime}$ for $i\in \cS^{(0)}$ means the abundance that the species $i$ would have if it had a non-zero component like the species of $\cS^{(+)}$.

\subsection{Eigenvalues of $H_{ij}^{(++)}$ and its largest one}

To obtain the eigenvalues $\lambda_i^{(+)}$'s of the $S^{(+)}\times S^{(+)}$ matrix $H_{ij}^{(++)}\equiv x_i^* B^{(+)}_{ij}$ at $x_i^*$, we decompose $H^{(++)}_{ij}$ as $H^{(++)}_{ij} = -x_i^*\delta_{ij} + V_{ij}$ with $V_{ij}\equiv - c x_i^* (1 - \delta_{ij}) + m x_i^* A_{ij}$. Considering  the eigenvalue expansion $\lambda_i^{(+)} \simeq  -x_i^* + V_{ii} + \sum_{j\neq i} {x_j^* V_{ji} V_{ij} x_i^* \over x_i^* -x_j^*}\simeq -x_i^*$ and noting that $V_{ii}=0$,  one  finds that the eigenvalues can be approximated by the zeroth-order term as $\lambda_i^{(+)} \simeq -x_i^*$ when $V_{ij}$  is sufficiently small. 

It should be also noted that $H_{ij}^{(++)}$ has $\vec{x}^* = (x_i^*)$ as an eigenvector with eigenvalue $-1$; $\sum_j H^{(++)}_{ij} x_j^* = \sum_j x_i^* B^{(+)}_{ij} \left(-\sum_\ell ((B^{(+)})^{-1})_{j\ell}\right) = - x_i^*$. Therefore the 
largest real part  of the eigenvalues $\lambda^{(+)}_i$'s is approximated by  $\max\left(-x_i^*, -1\right)$.

\section{$\bbB^{-1}$ to the first order in $m$}
\label{sec:firstorder}

For $\bbB = \bbB_0 + m \bbA$, one can expand its inverse $\bbB^{-1}$ in terms of the mutualism strength $m$ as 
\begin{align}
\bbB^{-1} &= \bbB_0^{-1} \sum_{n=0}^\infty \left( -m\bbA \bbB_0^{-1}\right)^n\nonumber\\
&= \bbB_0^{-1} - m \bbB_0^{-1} \bbA \bbB_0^{-1} + O(m^2).
\label{eq:BinvSeries}
\end{align}
Using Eqs.~(\ref{eq:AJ}), (\ref{eq:KJ}), and (\ref{eq:Binv0}), one can evaluate $\bbB^{-1}$ up to the first order of $m$ as 
\begin{align}
&\bbB^{-1}\simeq \bbB_0^{-1} - {m\over (1-c)^2} \left[-\bbI + \bbtc \bbtJ^{(0)}\right]\bbA\left[-\bbI + \bbtc \bbtJ^{(0)}\right]\nonumber\\
&=\bbB_0^{-1} - {m \over (1-c)^2} \left[ \bbA - \bbA \bbtc \bbtJ^{(0)} - \bbtc \bbtJ^{(0)}\bbA + \bbtc \bbtJ^{(0)} \bbA \bbtc \bbtJ^{(0)}\right]\nonumber\\
&=\bbB_0^{-1} - {m \over (1-c)^2} \left[ \bbA -\bbK \bbtN \bbtJ^{(1)} \bbtc - \bbtc \bbtJ^{(1)} \bbtN \left(\bbK - \langle \bbK \rangle \bbtc \right)\right].
\end{align}
Then the abundance of a plant species $p$ is given by 
\begin{align}
&x_p^{*(1)} = - \sum_j (B^{-1})_{pj}  =x_0^{\rm (P)} \nonumber\\
& + {m \over (1-c)^2} \left[k_p - k_p \tcA - \tcP {L\over \NP} + \tcP {L\over \NP} \tcA\right]\nonumber\\
&=x_0^{\rm (P)} + {m\over (1-c)^2} (1-\tcA) \left(k_p - \tcP \langle k\rangle^{\rm (P)}\right)\nonumber\\
&= x_0^{\rm (P)} \left[1 + {m \over 1-c} {1-\tcA \over 1-\tcP} \left(k_p - \tcP \langle k\rangle^{\rm (P)}\right) \right], 
\end{align}
and that of an animal species $a$ is 
\begin{align}
&x_a^{*(1)} =  x_0^{\rm (A)} \left[1 + {m \over 1-c} {1-\tcP \over 1-\tcA} \left(k_a - \tcA \langle k\rangle^{\rm (A)}\right) \right].
\end{align}
 
\section{$\bbB^{-1}$ under the annealed approximation for the adjacency matrix}
\label{sec:annealed}

In the annealed approximation, the adjacency matrix element $A_{ij}$ is approximated by the probability that the two nodes $i$ and $j$ are connected by a link in the ensemble of networks preserving the given degree sequence $\{k_i\}$ as 
\begin{equation}
\tilde{A}_{ij} = {k_i k_j\over L}
\end{equation} 
with $L$ the total number of links. Equivalently, the block adjacency matrix takes the form
\begin{equation}
\bbtA = {1\over L} \bbK \bbJ^{(1)} \bbK = {1\over \sqrt{\langle k\rangle^{\rm (P)} \langle k\rangle^{\rm (A)}} } \bbK \bbtJ^{(1)} \bbK.
\end{equation}
Using this, one finds that  the terms $\bbB_0^{-1} (-m \bbtA \bbB_0^{-1})^n$ appearing in Eq.~(\ref{eq:BinvExpand}) are simplified. Evaluating the first three terms, one can find the expression for general $n$ by induction. Let us first consider the term with $n=1$, which is evaluated as 
\begin{align}
&\bbB_0^{-1} (-m\bbtA) \bbB_0^{-1} \nonumber\\
&= {-m\over (1-c)^2} {\left(-\bbI + \bbtc \bbtJ^{(0)}\right) \bbK \bbtJ^{(1)} \bbK \left(-\bbI + \bbtc \bbtJ^{(0)}\right)\over \sqrt{\langle k\rangle^{\rm (P)} \langle k\rangle^{\rm (A)}}} \nonumber\\
&= {-m\over (1-c)^2} \times \nonumber\\
& {\bbK \bbtJ^{(1)}\bbK - \bbtc \langle \bbK\rangle \bbtJ^{(1)} \bbK - \bbK \bbtJ^{(1)} \bbtc \langle \bbK \rangle + \bbtc \langle \bbK\rangle \bbtJ^{(1)} \langle \bbK\rangle \bbtc \over \sqrt{\langle k\rangle^{\rm (P)} \langle k\rangle^{\rm (A)}}}\nonumber\\
&={-m\over (1-c)^2} {\left(\bbK - \bbtc \langle \bbK\rangle\right) \bbtJ^{(1)} \left(\bbK - \bbtc \langle \bbK\rangle\right)\over \sqrt{\langle k\rangle^{\rm (P)} \langle k\rangle^{\rm (A)}}}\nonumber\\
&=-{\tm \over 1-c} \bbtK\bbtJ^{(1)}\bbtK,
\label{eq:n1}
\end{align}
where Eq.~(\ref{eq:fJ}) is used and the rescaled degree matrix is introduced,
\begin{equation}
\bbtK \equiv {\bbK - \bbtc \langle \bbK \rangle \over \sqrt{\langle \bbK(\bbK  - \bbtc \langle \bbK\rangle)\rangle}} ={\bbK - \bbtc \langle \bbK \rangle \over \langle \bbtK \bbK\rangle},
\label{eq:Ktilde}
\end{equation}
and the rescaled mutualism strength $\tm$, defined in Eq.~(\ref{eq:mtilde}) in the main text, is here also represented as 
\begin{align}
\tm &= {m \over 1-c} {\langle \bbK \bbtK\rangle^{\rm (P)}\langle \bbK \bbtK\rangle^{\rm (A)} \over \sqrt{\langle \bbK\rangle^{\rm (P)}\langle \bbK\rangle^{\rm (A)}}}.
\end{align}
Notice that $\langle k\rangle^{\rm (P)} = \langle \bbK\rangle^{\rm (P)}$  and $\langle \bbK\bbtK\rangle^{\rm (P)} = \langle k^2\rangle^{\rm (P)} - \tcP (\langle k\rangle^{\rm (P)})^2 = (\langle k\rangle^{\rm (P)})^2 \left(\xiP - \tcP\right)$ from Eq.~(\ref{eq:fullave}). 

The term with $n=2$ is evaluated as 
\begin{align}
&\bbB_0^{-1} (-m\bbtA) \bbB_0^{-1} (-m\bbtA) \bbB_0^{-1} \nonumber\\
&={-\tm \over 1-c} \bbtK \bbtJ^{(1)} \bbtK  {-m\over \sqrt{\langle k\rangle^{\rm (P)} \langle k\rangle^{\rm (A)}}} \bbK \bbtJ^{(1)} \bbK {-\bbI + \bbtc \bbtJ^{(0)} \over 1-c}\nonumber\\
&={\tm \over 1-c} \bbtK {m \langle \bbtK \bbK\rangle^{\rm (P)} \langle \bbtK \bbK\rangle^{\rm (A)} \langle \bbtK \bbK\rangle^{-1} \bbtJ^{(0)}\over \sqrt{\langle k\rangle^{\rm (P)} \langle k\rangle^{\rm (A)}}}  {-\bbK + \bbtc \langle \bbK\rangle \over 1-c}\nonumber\\
&=- {\tm^2 \over 1-c} \bbtK \bbtJ^{(0)} \bbtK.
\label{eq:n2}
\end{align}

The term with $n=3$ is 
\begin{align}
&\bbB_0^{-1} (-m\bbtA) \bbB_0^{-1} (-m\bbtA) \bbB_0^{-1}(-m\bbtA) \bbB_0^{-1} \nonumber\\
&={-\tm^2 \over 1-c} \bbtK \bbtJ^{(0)} \bbtK  {-m\over \sqrt{\langle k\rangle^{\rm (P)} \langle k\rangle^{\rm (A)}}} \bbK \bbtJ^{(1)} \bbK {-\bbI + \bbtc \bbtJ^{(0)} \over 1-c}\nonumber\\
&={\tm^2 \over 1-c} \bbtK {m  \langle \bbtK \bbK\rangle \bbtJ^{(1)}\over \sqrt{\langle k\rangle^{\rm (P)} \langle k\rangle^{\rm (A)}}}  {-\bbK + \bbtc \langle \bbK\rangle \over 1-c}\nonumber\\
&=-  {\tm^2 \over 1-c}\bbtK  {m \langle \bbtK \bbK\rangle^{\rm (P)} \langle \bbtK \bbK\rangle^{\rm (A)} \over (1-c) \sqrt{\langle k\rangle^{\rm (P)} \langle k\rangle^{\rm (A)}}} \bbtJ^{(1)}\langle \bbK \bbtK\rangle^{-1} \left(\bbK - \bbtc \langle \bbK\rangle\right) \nonumber\\
&=- {\tm^3 \over 1-c} \bbtK \bbtJ^{(1)} \bbtK.
\label{eq:n3}
\end{align}

From Eqs.~(\ref{eq:n1}), (\ref{eq:n2}), and (\ref{eq:n3}), one can obtain by induction Eq.~(\ref{eq:Bn})
and we find that the inverse of the interaction matrix is evaluated as
\begin{align}
&\bbtB^{-1} = \bbB_0^{-1} -\nonumber\\
& {1\over 1-c} \bbtK \left(\bbtJ^{(1)}\sum_{n=1,3,5,\ldots}\tm^n + \bbtJ^{(0)} \sum_{n=2,4,6,\ldots} \tm^n \right)\bbtK  \nonumber\\
&=\bbB_0^{-1} - {1\over 1-c} {\tm \over 1-\tm^2} \bbtK \left(\tm \bbtJ^{(0)} + \bbtJ^{(1)}\right)\bbtK,
\label{eq:Binvdetails}
\end{align}
which is given also in Eq.~(\ref{eq:Binv}).

\section{Different measures of the species abundance in the long-time limit}
\label{sec:fcappendix}

\begin{figure*}
\includegraphics[width=2\columnwidth]{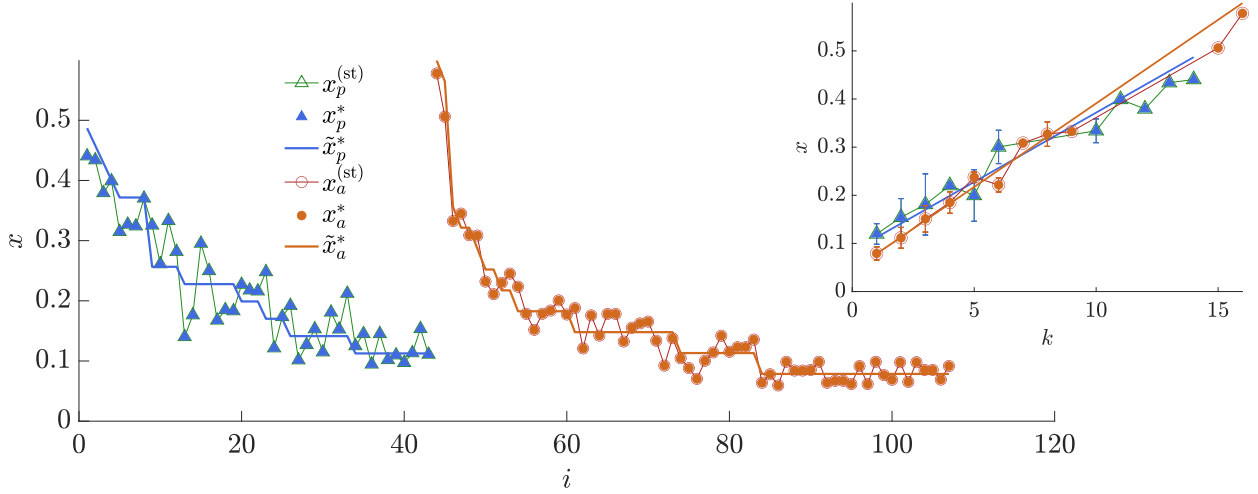}
\caption{Species abundance in the full coexistence phase. Stationary-state abundances $x_i^{\rm (st)}$ of individual species $i$ are compared with the stable fixed-point ones $x_i^*$'s and $\tx_i^*$'s for $c=0.1$ and $\tm=0.4$. The species index $i$ is arranged in the descending order of degree among plants and among animals. Inset: Abundance vs. degree.}
\label{fig:tm0p4}
\end{figure*}

In the present study appear a couple of different measures of the species abundance, being numerical or  analytical solutions to the LV equations. We summarize their notations here to help distinguish and understand them.
\begin{itemize}
\item $x_i^{\rm (st)}$ : It is defined in Eq.~(\ref{eq:xist}) and represents the numerical solution $x_i(T)$ at the final step to the LV equation in Eq.~(\ref{eq:LV}) with using the interaction matrix $\bbB$ constructed from the original adjacency matrix $\bbA$. 
\item $\tx_i^{\rm (st)}$ : It represents the numerical solution $\tx_i(T)$ at the final step to the LV equation in Eq.~(\ref{eq:LV}) with using the interaction matrix $\bbtB$ constructed from the factorized adjacency matrix $\bbtA$ in Eq.~(\ref{eq:Atilde}) under the annealed approximation.
\item $x_i^*$ : It is defined in Eq.~(\ref{eq:xi*}) and represents the stable fixed point to the LV equation in Eq.~(\ref{eq:LV}) with using $\bbB$. The sets of surviving and extinct species, $\cS^{(+)}$ and $\cS^{(0)}$ can be obtained by updating iteratively $\bbB^{(+)}$ and $x_i^*$, using Eq.~(\ref{eq:xi*}), as described in Sec.~\ref{sec:sephase}.  
\item $\tx_i^*$ : It represents the stable fixed point to the LV equation in Eq.~(\ref{eq:LV}) with using $\bbtB$ under the annealed approximation. While it can be evaluated numerically by Eq.~(\ref{eq:xi*}) with using $\bbtB$, its analytic expression is available in Eq.~(\ref{eq:xfpfulleff}). The sets of surviving and extinct species, $\cS^{(+)}$ and $\cS^{(0)}$, can be obtained by updating iteratively $\bbtB^{(+)}$ and $\tx_i^*$, using Eq.~(\ref{eq:xfpfulleff}), as described in Sec.~\ref{sec:sephase}. 
\end{itemize}

In Fig.~\ref{fig:tm0p4}, we present $x_i^{\rm (st)}$, $x_i^*$ and $\tx_i^*$  for  $c=0.1$ and $\tm=0.4$ in the full coexistence phase. 

\section{Effective quantities in Eq.~(\ref{eq:xfpfulleff})}
\label{sec:effective}

\begin{figure}[h]
\includegraphics[width=\columnwidth]{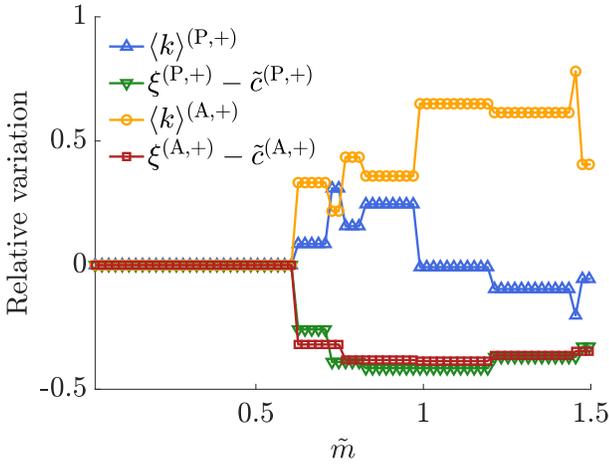}
\caption{Relative variations of the network structural properties with  the rescaled mutualism strength $\tm$ for $c=0.1$.  Shown are the relative variations of the mean degree $\langle k\rangle$ and the degree heterogeneity minus the rescaled competition $\xi - \tilde{c}$ for plants and animals.  The relative variation is evaluated e.g., as ${\langle k\rangle^{\rm (P,+)} - \langle k\rangle^{\rm (P)} \over \langle k\rangle^{\rm (P)}}$. }
\label{fig:variationeff}
\end{figure}

The effective interaction matrix $\bbtB^{(+)}$ for the surviving species is obtained by removing the rows and columns corresponding to the species belonging to $\cS^{(0)}$ in the full interaction matrix $\bbtB$. For later use, we introduce $\cS^{\rm (P,+)}$ and $\cS^{\rm (A,+)}$ to denote the set of  plant and animal species, respectively, in $\cS^{(+)}$ to be assigned non-zero components, and $\cS^{\rm (P,0)}$ and $\cS^{\rm (A,0)}$ to denote the set of plant and animal species, respectively, in $\cS^{(0)}$ to be assigned zero components. Their sizes are $N^{\rm (P,+)}$, $N^{\rm (A,+)}$, $N^{\rm (P,0)}$, and $N^{\rm (A,0)}$. The effective interaction matrix $\bbtB^{(+)}$ is of size $S^{(+)}\times S^{(+)}$ with $S^{(+)} = N^{\rm (P,+)}+N^{\rm (A,+)}$ and takes the form 
\begin{equation}
\bbtB^{(+)} = -\bbI^{(+)} + c(\bbJ^{\rm (0,+)} - \bbI^{(+)}) + m\bbtA^{(+)},
\end{equation}
where  $\bbtA^{(+)}$ is the effective adjacency matrix for the plant and animal species in $\cS^{(+)}$.

The effective  adjacency matrix $\bbtA^{(+)}= {\bf \tA}^{\rm (PA,+)} \oplus {\bf \tA}^{\rm (AP,+)}$ is  obtained by removing in $\bbtA$ the rows and columns of the species in $\cS^{(0)}$. Therefore it holds that $\tA^{(+)}_{pa} = \tA_{pa}= k_p k_a/L$ if $p$ and $a$ are in $\cS^{(+)}$.  Once $\bbtA^{(+)}$ is  given,  the effective network quantities such as the effective degree can be derived from $\bbtA^{(+)}$. Moreover, $\bbtA^{(+)}$ maintains the factorized form in terms of the effective degrees and the effective number of links. Therefore the effective quantities can be inserted into Eq.~(\ref{eq:xfpfull}), developed originally with the factorized adjacency matrix, to yield Eq.~(\ref{eq:xfpfulleff}). Below we present how to evaluate  them specifically.

(i) The effective rescaled competition strength is $\tilde{c}^{\rm (G,+)} \equiv  {c N ^{\rm (G,+)} \over c N^{\rm (G,+)} + 1-c}$ with G being P or A.

(ii) The effective zeroth-order abundance is $x_0^{\rm (G, +)} \equiv {1- \tilde{c}^{\rm (G, +)} \over 1-c}$.

(iii)  The total number of links is $L^{(+)}\equiv\sum_{p\in \cS^{\rm (P,+)}, a\in \cS^{\rm (A,+)}} \tA^{(+)}_{pa} = {1\over L} \sum_{p\in \cS^{\rm (P,+)}} k_p \sum_{a\in \cS^{\rm (A,+)}} k_a = L \,\ell^{\rm (P,+)} \ell^{\rm (A,+)}$ with  $\ell^{\rm (P,+)} \equiv {1\over L} \sum_{p\in \cS^{\rm (P,+)}} k_p$ and $\ell^{\rm (A,+)}\equiv{1\over L} \sum_{a\in \cS^{\rm (A,+)}} k_a$ denoting the ratio of the links incident on the  plant and animal species of $\cS^{(+)}$, respectively, to the original number of links $L = \sum_p k_p = \sum_a k_a$. 

(iv) The effective degree of a plant or animal species in $\cS^{(+)}$  is evaluated as $k_p^{(+)} \equiv \sum_{a\in \cS^{\rm (A,+)}} \tA^{(+)}_{pa}= k_p \sum_{a\in \cS^{\rm (A,+)} }{k_a \over L} = k_p \ell^{\rm (A,+)}$ and $k_a^{(+)}\equiv\sum_{p\in \cS^{\rm (P,+)}} \tA^{(+)}_{pa}= k_a \ell^{\rm (P,+)}$, satisfying $L^{(+)} = \sum_{p\in \cS^{\rm (P,+)}} k_p^{(+)}= \sum_{a\in \cS^{\rm (A,+)}} k_a^{(+)}$.

(v) The effective adjacency matrix maintains its factorized form $\tA^{(+)}_{pa} = \tA_{pa}= {k_p k_a \over L} = {k_p^{(+)} k_a^{(+)} \over L^{(+)}}$ in terms of the effective degrees and the effective numbers of links defined above.

(vi) The effective degree heterogeneity is evaluated as $\xi^{\rm (P,+)} = {\langle k^2\rangle^{\rm (P,+)} \over \langle k\rangle^{\rm (P,+)}}$ with the  moments given by $\langle k^n\rangle^{\rm (P,+)} = {1\over N^{\rm (P,+)}} \sum_{p\in \cS^{\rm (P,+)}} \left(k_p^{(+)}\right)^n$. $\xi^{\rm (A,+)}$ and $\langle k^n\rangle^{\rm (A,+)}$ are evaluated in the same manner. 

(vii) The effective rescaled degree is evaluated by $\bk_i^{(+)} \equiv {{k_i^{(+)} \over \langle k\rangle^{({\rm G}_i,+)}}-\tilde{c}^{({\rm G}_i,+)} \over \xi^{({\rm G}_i,+)} - \tilde{c}^{({\rm G}_i,+)}}$.

(viii) The effective asymmetry factor is evaluated by $\eta^{\rm (PA,+)} \equiv {1-\tilde{c}^{\rm (A,+)} \over 1-\tilde{c}^{\rm (P,+)}} \sqrt{\langle k\rangle^{\rm (P,+)} (\xi^{\rm (P,+)} - \tilde{c}^{\rm (P,+)}) \over \langle k\rangle^{\rm (A,+)} (\xi^{\rm (A,+)}- \tilde{c}^{\rm (A,+)})} = {1\over \eta^{\rm (AP,+)}}$.

(ix) The effective rescaled mutualism strength $\tm^{(+)}$  is evaluated by 
\begin{align}
\tilde{m}^{(+)}\equiv& \frac{m}{1-c}  \nonumber\\
\times&\sqrt{\langle k\rangle^{(P,+)} \langle k\rangle^{(A,+)} (\xi^{(P,+)}-\tilde{c}^{(P,+)}) (\xi^{(A,+)}-\tilde{c}^{(A,+)})}
\label{eq:mtildedetails}
\end{align} 
exactly in the same manner as Eq. (\ref{eq:mtilde}) with using the effective quantities. 

The relative variations of the effective quantities with respect to their original values are shown in Fig.~\ref{fig:variationeff}. While the extinction of small-degree species may make the effective mean degree $\langle k\rangle^{(+)}$  larger than the original one $\langle k\rangle$, the hub plants and animals lose their significant portions of partners, resulting in the reduction of the effective degree heterogeneity. The rescaled competition $\tilde{c}^{(+)}$ decreases as more species go extinct with increasing $\tm$. These quantities together determine the effective rescaled mutualism $\tm^{(+)}$, which turns out to be smaller than $\tm$ as shown in Fig.~\ref{fig:extinction} (b). 

\section{Species abundance under the annealed adjacency matrix}
\label{sec:abundanceannealed}


\begin{figure*}
\includegraphics[width=2\columnwidth]{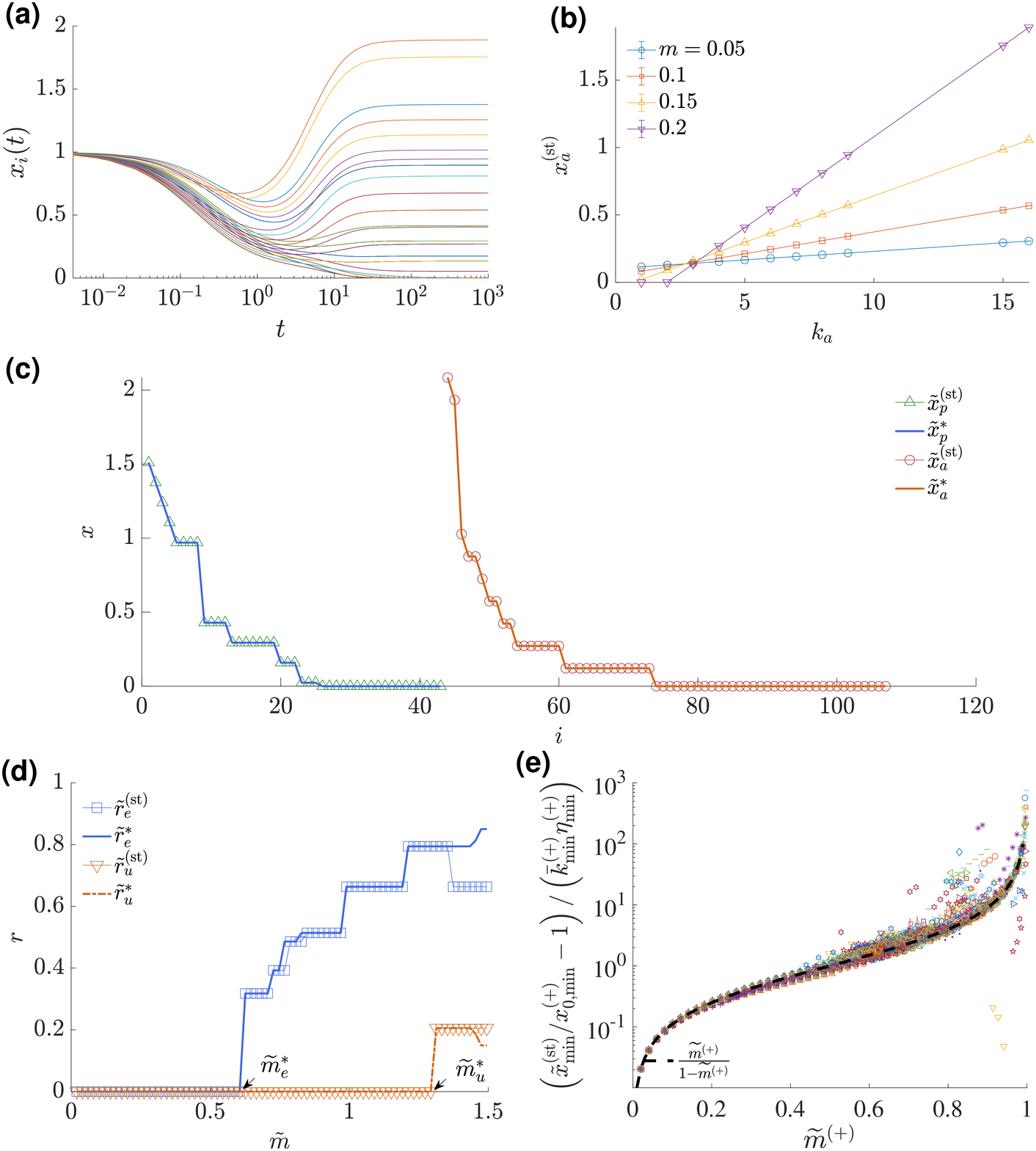}
\caption{Species abundance under the annealed adjacency matrix. 
(a) Time-evolution of the abundances of individual species  (different lines) obtained by numerically integrating Eq.~(\ref{eq:LV}) with the annealed adjacency matrix $\tilde{A}_{ij}$ used for $c=0.1$ and $m=0.2$. 
(b) The stationary-state abundance vs.  degree for animal species  with $c=0.1$.
(c)  Stationary-state abundances $\tx_i^{\rm (st)}$ of individual species $i$ are compared with the stable fixed point components $\tx_i^*$'s for $c=0.1$ and $\tm=0.8$ belonging to the selective extinction phase. 
(d) Fraction $\tr_e^{\rm (st)}$ of extinct species and $\tr_u$ of the abundance-diverging species based on the stationary-state abundance $\tx_i^{\rm (st)}$. They are compared with the theoretical predictions $\tr_e^*$ and  $\tr_u^*$ based on the stable fixed point $\tx_i^*$'s.     
(e) The collapse of the minimum abundances $\tx_{\rm min}^{\rm (st)}$ rescaled as  in Eq.~(\ref{eq:scal}) in 46 real-world communities as functions of  $\tm^{(+)}$.
}
\label{fig:Eff_annealed}
\end{figure*}

Here we present the plots of the species abundances, the fraction of extinct and abundance-diverging species, and the rescaled minimum abundance in case of the annealed interaction matrix $\bbtB$ and the annealed adjacency matrix $\bbtA$ in Fig.~\ref{fig:Eff_annealed}.

\section{Accuracy in the prediction of the extinction of individual species}
\label{sec:accuracy}

\begin{figure}
\includegraphics[width=\columnwidth]{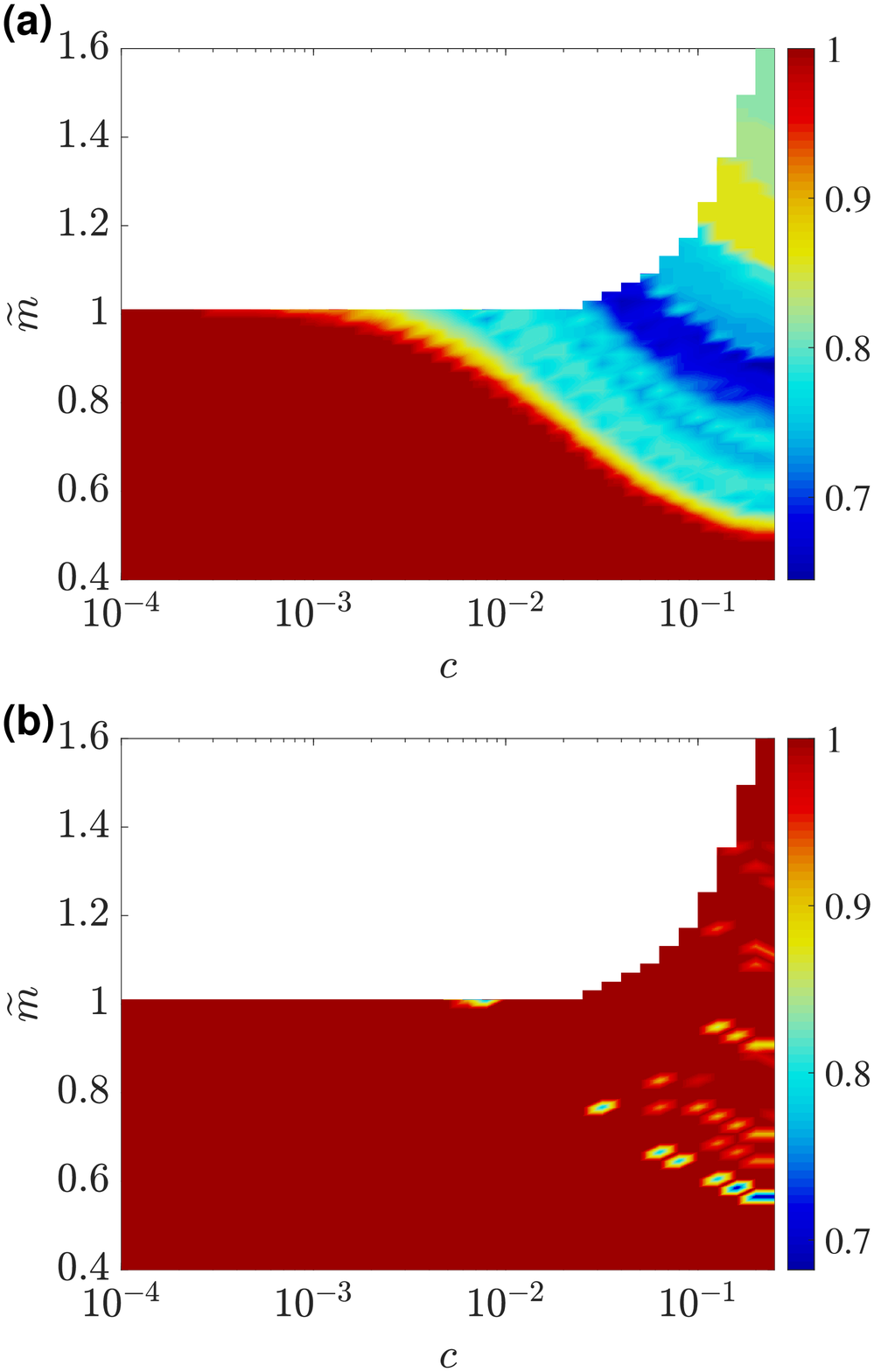}
\caption{Accuracy of the analytic formula of the stable fixed-point abundances in predicting the extinction or survival of individual species. Shown in the $(c,\tm)$ plane is the  fraction of the species whose survival or extinction is predicted identically by (a) both the stationary-state abundance $x_i^{\rm (st)}$ under the original interaction matrix $\bbB$, and the stable fixed point $\tx_i^*$,  and (b) both $\tx_i^{\rm (st)}$  under the annealed interaction matrix  $\bbtB$, and the stable fixed point $\tx_i^*$.}
\label{fig:accuracy}
\end{figure}

We consider a species extinct if its abundance  is smaller than $\epsilon=10^{-5}$  and surviving otherwise. The criterion is used to assess the stationary-state abundance $x_i^{\rm (st)}$ and $\tx_i^{\rm (st)}$ and discriminate the fate of $i$ evolving under the original and the annealed interaction matrix, respectively. To illuminate the predictive power of the stable fixed point abundance $\tx_i^{*}$ for the fate - survival or extinction - of individual species, we compute the fraction of the species that are correctly predicted, i.e., found to be surviving in both abundances, $x_i^{\rm (st)}\geq \epsilon$ and $\tx_i^*\geq \epsilon$ or found to be extinct in both, $x_i^{\rm (st)}<\epsilon$ and $\tx_i^*< \epsilon$, which we can consider as the accuracy of the stable  fixed point abundances in the prediction of species extinction and present in Fig.~\ref{fig:accuracy} (a).  We also do the same analysis with $\tx_i^{\rm (st)}$ and $\tx_i^*$ and show the result in Fig.~\ref{fig:accuracy} (b).  On the average across parameters  in the selective extinction phase, the accuracy of $\tx_i^*$ in predicting extinction/survival amounts to 79.3\% for $x_i^{\rm (st)}$ under the original interaction matrix and  99.2\% for $\tx_i^{\rm (st)}$ under the annealed interaction matrix.

\bibliographystyle{apsrev4-1}
\bibliography{references}

\begin{thebibliography}{38}%
\makeatletter
\providecommand \@ifxundefined [1]{%
 \@ifx{#1\undefined}
}%
\providecommand \@ifnum [1]{%
 \ifnum #1\expandafter \@firstoftwo
 \else \expandafter \@secondoftwo
 \fi
}%
\providecommand \@ifx [1]{%
 \ifx #1\expandafter \@firstoftwo
 \else \expandafter \@secondoftwo
 \fi
}%
\providecommand \natexlab [1]{#1}%
\providecommand \enquote  [1]{``#1''}%
\providecommand \bibnamefont  [1]{#1}%
\providecommand \bibfnamefont [1]{#1}%
\providecommand \citenamefont [1]{#1}%
\providecommand \href@noop [0]{\@secondoftwo}%
\providecommand \href [0]{\begingroup \@sanitize@url \@href}%
\providecommand \@href[1]{\@@startlink{#1}\@@href}%
\providecommand \@@href[1]{\endgroup#1\@@endlink}%
\providecommand \@sanitize@url [0]{\catcode `\\12\catcode `\$12\catcode
  `\&12\catcode `\#12\catcode `\^12\catcode `\_12\catcode `\%12\relax}%
\providecommand \@@startlink[1]{}%
\providecommand \@@endlink[0]{}%
\providecommand \url  [0]{\begingroup\@sanitize@url \@url }%
\providecommand \@url [1]{\endgroup\@href {#1}{\urlprefix }}%
\providecommand \urlprefix  [0]{URL }%
\providecommand \Eprint [0]{\href }%
\providecommand \doibase [0]{http://dx.doi.org/}%
\providecommand \selectlanguage [0]{\@gobble}%
\providecommand \bibinfo  [0]{\@secondoftwo}%
\providecommand \bibfield  [0]{\@secondoftwo}%
\providecommand \translation [1]{[#1]}%
\providecommand \BibitemOpen [0]{}%
\providecommand \bibitemStop [0]{}%
\providecommand \bibitemNoStop [0]{.\EOS\space}%
\providecommand \EOS [0]{\spacefactor3000\relax}%
\providecommand \BibitemShut  [1]{\csname bibitem#1\endcsname}%
\let\auto@bib@innerbib\@empty
\bibitem [{\citenamefont {May}(1972)}]{MAY:1972aa}%
  \BibitemOpen
  \bibfield  {author} {\bibinfo {author} {\bibfnamefont {R.~M.}\ \bibnamefont
  {May}},\ }\href {http://dx.doi.org/10.1038/238413a0} {\bibfield  {journal}
  {\bibinfo  {journal} {Nature}\ }\textbf {\bibinfo {volume} {238}},\ \bibinfo
  {pages} {413} (\bibinfo {year} {1972})}\BibitemShut {NoStop}%
\bibitem [{\citenamefont {Goh}(1979)}]{Goh1979}%
  \BibitemOpen
  \bibfield  {author} {\bibinfo {author} {\bibfnamefont {B.~S.}\ \bibnamefont
  {Goh}},\ }\href {\doibase 10.1086/283384} {\bibfield  {journal} {\bibinfo
  {journal} {Am. Nat.}\ }\textbf {\bibinfo {volume} {113}},\ \bibinfo {pages}
  {261} (\bibinfo {year} {1979})}\BibitemShut {NoStop}%
\bibitem [{\citenamefont {Allesina}\ and\ \citenamefont
  {Tang}(2015)}]{Allesina2015}%
  \BibitemOpen
  \bibfield  {author} {\bibinfo {author} {\bibfnamefont {S.}~\bibnamefont
  {Allesina}}\ and\ \bibinfo {author} {\bibfnamefont {S.}~\bibnamefont
  {Tang}},\ }\href {\doibase 10.1007/s10144-014-0471-0} {\bibfield  {journal}
  {\bibinfo  {journal} {Popul. Ecol.}\ }\textbf {\bibinfo {volume} {57}},\
  \bibinfo {pages} {63} (\bibinfo {year} {2015})}\BibitemShut {NoStop}%
\bibitem [{\citenamefont {Stone}(2016)}]{Stone2016}%
  \BibitemOpen
  \bibfield  {author} {\bibinfo {author} {\bibfnamefont {L.}~\bibnamefont
  {Stone}},\ }\href {\doibase 10.1038/ncomms12857} {\bibfield  {journal}
  {\bibinfo  {journal} {Nat. Commun.}\ }\textbf {\bibinfo {volume} {7}},\
  \bibinfo {pages} {1} (\bibinfo {year} {2016})}\BibitemShut {NoStop}%
\bibitem [{\citenamefont {Grilli}\ \emph {et~al.}(2017)\citenamefont {Grilli},
  \citenamefont {Adorisio}, \citenamefont {Suweis}, \citenamefont
  {Barab{\'a}s}, \citenamefont {Banavar}, \citenamefont {Allesina},\ and\
  \citenamefont {Maritan}}]{Grilli:2017wj}%
  \BibitemOpen
  \bibfield  {author} {\bibinfo {author} {\bibfnamefont {J.}~\bibnamefont
  {Grilli}}, \bibinfo {author} {\bibfnamefont {M.}~\bibnamefont {Adorisio}},
  \bibinfo {author} {\bibfnamefont {S.}~\bibnamefont {Suweis}}, \bibinfo
  {author} {\bibfnamefont {G.}~\bibnamefont {Barab{\'a}s}}, \bibinfo {author}
  {\bibfnamefont {J.~R.}\ \bibnamefont {Banavar}}, \bibinfo {author}
  {\bibfnamefont {S.}~\bibnamefont {Allesina}}, \ and\ \bibinfo {author}
  {\bibfnamefont {A.}~\bibnamefont {Maritan}},\ }\href {\doibase
  10.1038/ncomms14389} {\bibfield  {journal} {\bibinfo  {journal} {Nat.
  Commun.}\ }\textbf {\bibinfo {volume} {8}},\ \bibinfo {pages} {14389}
  (\bibinfo {year} {2017})}\BibitemShut {NoStop}%
\bibitem [{\citenamefont {Bunin}(2017)}]{PhysRevE.95.042414}%
  \BibitemOpen
  \bibfield  {author} {\bibinfo {author} {\bibfnamefont {G.}~\bibnamefont
  {Bunin}},\ }\href {\doibase 10.1103/PhysRevE.95.042414} {\bibfield  {journal}
  {\bibinfo  {journal} {Phys. Rev. E}\ }\textbf {\bibinfo {volume} {95}},\
  \bibinfo {pages} {042414} (\bibinfo {year} {2017})}\BibitemShut {NoStop}%
\bibitem [{\citenamefont {Cui}\ \emph {et~al.}(2020)\citenamefont {Cui},
  \citenamefont {Marsland},\ and\ \citenamefont
  {Mehta}}]{PhysRevLett.125.048101}%
  \BibitemOpen
  \bibfield  {author} {\bibinfo {author} {\bibfnamefont {W.}~\bibnamefont
  {Cui}}, \bibinfo {author} {\bibfnamefont {R.}~\bibnamefont {Marsland}}, \
  and\ \bibinfo {author} {\bibfnamefont {P.}~\bibnamefont {Mehta}},\ }\href
  {\doibase 10.1103/PhysRevLett.125.048101} {\bibfield  {journal} {\bibinfo
  {journal} {Phys. Rev. Lett.}\ }\textbf {\bibinfo {volume} {125}},\ \bibinfo
  {pages} {048101} (\bibinfo {year} {2020})}\BibitemShut {NoStop}%
\bibitem [{\citenamefont {Pettersson}\ \emph
  {et~al.}(2020{\natexlab{a}})\citenamefont {Pettersson}, \citenamefont
  {Savage},\ and\ \citenamefont {Jacobi}}]{doi:10.1098/rsif.2019.0391}%
  \BibitemOpen
  \bibfield  {author} {\bibinfo {author} {\bibfnamefont {S.}~\bibnamefont
  {Pettersson}}, \bibinfo {author} {\bibfnamefont {V.~M.}\ \bibnamefont
  {Savage}}, \ and\ \bibinfo {author} {\bibfnamefont {M.~N.}\ \bibnamefont
  {Jacobi}},\ }\href {\doibase 10.1098/rsif.2019.0391} {\bibfield  {journal}
  {\bibinfo  {journal} {J. R. Soc. Interface}\ }\textbf {\bibinfo {volume}
  {17}},\ \bibinfo {pages} {20190391} (\bibinfo {year}
  {2020}{\natexlab{a}})}\BibitemShut {NoStop}%
\bibitem [{\citenamefont {Pettersson}\ \emph
  {et~al.}(2020{\natexlab{b}})\citenamefont {Pettersson}, \citenamefont
  {Savage},\ and\ \citenamefont {Jacobi}}]{Pettersson2020}%
  \BibitemOpen
  \bibfield  {author} {\bibinfo {author} {\bibfnamefont {S.}~\bibnamefont
  {Pettersson}}, \bibinfo {author} {\bibfnamefont {V.~M.}\ \bibnamefont
  {Savage}}, \ and\ \bibinfo {author} {\bibfnamefont {M.~N.}\ \bibnamefont
  {Jacobi}},\ }\href {\doibase 10.1103/PhysRevE.102.062405} {\bibfield
  {journal} {\bibinfo  {journal} {Phys. Rev. E}\ }\textbf {\bibinfo {volume}
  {102}},\ \bibinfo {pages} {062405} (\bibinfo {year}
  {2020}{\natexlab{b}})}\BibitemShut {NoStop}%
\bibitem [{\citenamefont {Montoya}\ and\ \citenamefont
  {Sol{\'e}}(2002)}]{MONTOYA2002405}%
  \BibitemOpen
  \bibfield  {author} {\bibinfo {author} {\bibfnamefont {J.~M.}\ \bibnamefont
  {Montoya}}\ and\ \bibinfo {author} {\bibfnamefont {R.~V.}\ \bibnamefont
  {Sol{\'e}}},\ }\href {\doibase http://dx.doi.org/10.1006/jtbi.2001.2460}
  {\bibfield  {journal} {\bibinfo  {journal} {J. Theor. Biol.}\ }\textbf
  {\bibinfo {volume} {214}},\ \bibinfo {pages} {405 } (\bibinfo {year}
  {2002})}\BibitemShut {NoStop}%
\bibitem [{\citenamefont {Dunne}\ \emph {et~al.}(2002)\citenamefont {Dunne},
  \citenamefont {Williams},\ and\ \citenamefont {Martinez}}]{Dunne2002}%
  \BibitemOpen
  \bibfield  {author} {\bibinfo {author} {\bibfnamefont {J.~A.}\ \bibnamefont
  {Dunne}}, \bibinfo {author} {\bibfnamefont {R.~J.}\ \bibnamefont {Williams}},
  \ and\ \bibinfo {author} {\bibfnamefont {N.~D.}\ \bibnamefont {Martinez}},\
  }\href {\doibase 10.1073/pnas.192407699} {\bibfield  {journal} {\bibinfo
  {journal} {Proc. Natl. Acad. Sci. U.S.A.}\ }\textbf {\bibinfo {volume}
  {99}},\ \bibinfo {pages} {12917} (\bibinfo {year} {2002})}\BibitemShut
  {NoStop}%
\bibitem [{\citenamefont {Bascompte}\ \emph {et~al.}(2003)\citenamefont
  {Bascompte}, \citenamefont {Jordano}, \citenamefont {Meli{\'{a}}n},\ and\
  \citenamefont {Olesen}}]{Bascompte2003}%
  \BibitemOpen
  \bibfield  {author} {\bibinfo {author} {\bibfnamefont {J.}~\bibnamefont
  {Bascompte}}, \bibinfo {author} {\bibfnamefont {P.}~\bibnamefont {Jordano}},
  \bibinfo {author} {\bibfnamefont {C.~J.}\ \bibnamefont {Meli{\'{a}}n}}, \
  and\ \bibinfo {author} {\bibfnamefont {J.~M.}\ \bibnamefont {Olesen}},\
  }\href {\doibase 10.1073/pnas.1633576100} {\bibfield  {journal} {\bibinfo
  {journal} {Proc. Natl. Acad. Sci. U.S.A.}\ }\textbf {\bibinfo {volume}
  {100}},\ \bibinfo {pages} {9383} (\bibinfo {year} {2003})}\BibitemShut
  {NoStop}%
\bibitem [{\citenamefont {Jordano}\ \emph {et~al.}(2003)\citenamefont
  {Jordano}, \citenamefont {Bascompte},\ and\ \citenamefont
  {Olesen}}]{Jordano2003}%
  \BibitemOpen
  \bibfield  {author} {\bibinfo {author} {\bibfnamefont {P.}~\bibnamefont
  {Jordano}}, \bibinfo {author} {\bibfnamefont {J.}~\bibnamefont {Bascompte}},
  \ and\ \bibinfo {author} {\bibfnamefont {J.~M.}\ \bibnamefont {Olesen}},\
  }\href {\doibase 10.1046/j.1461-0248.2003.00403.x} {\bibfield  {journal}
  {\bibinfo  {journal} {Ecol. Lett.}\ }\textbf {\bibinfo {volume} {6}},\
  \bibinfo {pages} {69} (\bibinfo {year} {2003})}\BibitemShut {NoStop}%
\bibitem [{\citenamefont {Montoya}\ \emph {et~al.}(2006)\citenamefont
  {Montoya}, \citenamefont {Pimm},\ and\ \citenamefont
  {Sol{\'{e}}}}]{Montoya2006}%
  \BibitemOpen
  \bibfield  {author} {\bibinfo {author} {\bibfnamefont {J.~M.}\ \bibnamefont
  {Montoya}}, \bibinfo {author} {\bibfnamefont {S.~L.}\ \bibnamefont {Pimm}}, \
  and\ \bibinfo {author} {\bibfnamefont {R.~V.}\ \bibnamefont {Sol{\'{e}}}},\
  }\href {\doibase 10.1038/nature04927} {\bibfield  {journal} {\bibinfo
  {journal} {Nature}\ }\textbf {\bibinfo {volume} {442}},\ \bibinfo {pages}
  {259} (\bibinfo {year} {2006})}\BibitemShut {NoStop}%
\bibitem [{\citenamefont {Bascompte}\ and\ \citenamefont
  {Jordano}(2007)}]{Bascompte2007}%
  \BibitemOpen
  \bibfield  {author} {\bibinfo {author} {\bibfnamefont {J.}~\bibnamefont
  {Bascompte}}\ and\ \bibinfo {author} {\bibfnamefont {P.}~\bibnamefont
  {Jordano}},\ }\href {\doibase 10.1146/annurev.ecolsys.38.091206.095818}
  {\bibfield  {journal} {\bibinfo  {journal} {Annu. Rev. Ecol. Evol. Syst.}\
  }\textbf {\bibinfo {volume} {38}},\ \bibinfo {pages} {567} (\bibinfo {year}
  {2007})}\BibitemShut {NoStop}%
\bibitem [{\citenamefont {Guimar{\~{a}}es}\ \emph {et~al.}(2007)\citenamefont
  {Guimar{\~{a}}es}, \citenamefont {Machado}, \citenamefont {de~Aguiar},
  \citenamefont {Jordano}, \citenamefont {Bascompte}, \citenamefont
  {Pinheiro},\ and\ \citenamefont {dos Reis}}]{Guimaraes2007}%
  \BibitemOpen
  \bibfield  {author} {\bibinfo {author} {\bibfnamefont {P.~R.}\ \bibnamefont
  {Guimar{\~{a}}es}}, \bibinfo {author} {\bibfnamefont {G.}~\bibnamefont
  {Machado}}, \bibinfo {author} {\bibfnamefont {M.~A.}\ \bibnamefont
  {de~Aguiar}}, \bibinfo {author} {\bibfnamefont {P.}~\bibnamefont {Jordano}},
  \bibinfo {author} {\bibfnamefont {J.}~\bibnamefont {Bascompte}}, \bibinfo
  {author} {\bibfnamefont {A.}~\bibnamefont {Pinheiro}}, \ and\ \bibinfo
  {author} {\bibfnamefont {S.~F.}\ \bibnamefont {dos Reis}},\ }\href {\doibase
  10.1016/j.jtbi.2007.08.004} {\bibfield  {journal} {\bibinfo  {journal} {J.
  Theor. Biol.}\ }\textbf {\bibinfo {volume} {249}},\ \bibinfo {pages} {181}
  (\bibinfo {year} {2007})}\BibitemShut {NoStop}%
\bibitem [{\citenamefont {Olesen}\ \emph {et~al.}(2007)\citenamefont {Olesen},
  \citenamefont {Bascompte}, \citenamefont {Dupont},\ and\ \citenamefont
  {Jordano}}]{Olesen2007}%
  \BibitemOpen
  \bibfield  {author} {\bibinfo {author} {\bibfnamefont {J.~M.}\ \bibnamefont
  {Olesen}}, \bibinfo {author} {\bibfnamefont {J.}~\bibnamefont {Bascompte}},
  \bibinfo {author} {\bibfnamefont {Y.~L.}\ \bibnamefont {Dupont}}, \ and\
  \bibinfo {author} {\bibfnamefont {P.}~\bibnamefont {Jordano}},\ }\href
  {\doibase 10.1073/pnas.0706375104} {\bibfield  {journal} {\bibinfo  {journal}
  {Proc. Natl. Acad. Sci. U.S.A.}\ }\textbf {\bibinfo {volume} {104}},\
  \bibinfo {pages} {19891} (\bibinfo {year} {2007})}\BibitemShut {NoStop}%
\bibitem [{\citenamefont {Th{\'e}bault}\ and\ \citenamefont
  {Fontaine}(2010)}]{thebault10}%
  \BibitemOpen
  \bibfield  {author} {\bibinfo {author} {\bibfnamefont {E.}~\bibnamefont
  {Th{\'e}bault}}\ and\ \bibinfo {author} {\bibfnamefont {C.}~\bibnamefont
  {Fontaine}},\ }\href@noop {} {\bibfield  {journal} {\bibinfo  {journal}
  {Science}\ }\textbf {\bibinfo {volume} {329}},\ \bibinfo {pages} {853}
  (\bibinfo {year} {2010})}\BibitemShut {NoStop}%
\bibitem [{\citenamefont {Maeng}\ and\ \citenamefont {Lee}(2011)}]{Maeng2011}%
  \BibitemOpen
  \bibfield  {author} {\bibinfo {author} {\bibfnamefont {S.~E.}\ \bibnamefont
  {Maeng}}\ and\ \bibinfo {author} {\bibfnamefont {J.~W.}\ \bibnamefont
  {Lee}},\ }\href {\doibase 10.3938/jkps.58.851} {\bibfield  {journal}
  {\bibinfo  {journal} {J. Korean Phys. Soc.}\ }\textbf {\bibinfo {volume}
  {58}},\ \bibinfo {pages} {851} (\bibinfo {year} {2011})}\BibitemShut
  {NoStop}%
\bibitem [{\citenamefont {Bastolla}\ \emph {et~al.}(2009)\citenamefont
  {Bastolla}, \citenamefont {Fortuna}, \citenamefont {Pascual-Garc{\'{i}}a},
  \citenamefont {Ferrera}, \citenamefont {Luque},\ and\ \citenamefont
  {Bascompte}}]{Bastolla2009}%
  \BibitemOpen
  \bibfield  {author} {\bibinfo {author} {\bibfnamefont {U.}~\bibnamefont
  {Bastolla}}, \bibinfo {author} {\bibfnamefont {M.~A.}\ \bibnamefont
  {Fortuna}}, \bibinfo {author} {\bibfnamefont {A.}~\bibnamefont
  {Pascual-Garc{\'{i}}a}}, \bibinfo {author} {\bibfnamefont {A.}~\bibnamefont
  {Ferrera}}, \bibinfo {author} {\bibfnamefont {B.}~\bibnamefont {Luque}}, \
  and\ \bibinfo {author} {\bibfnamefont {J.}~\bibnamefont {Bascompte}},\ }\href
  {\doibase 10.1038/nature07950} {\bibfield  {journal} {\bibinfo  {journal}
  {Nature}\ }\textbf {\bibinfo {volume} {458}},\ \bibinfo {pages} {1018}
  (\bibinfo {year} {2009})}\BibitemShut {NoStop}%
\bibitem [{\citenamefont {Suweis}\ \emph {et~al.}(2013)\citenamefont {Suweis},
  \citenamefont {Simini}, \citenamefont {Banavar},\ and\ \citenamefont
  {Maritan}}]{suweis2013}%
  \BibitemOpen
  \bibfield  {author} {\bibinfo {author} {\bibfnamefont {S.}~\bibnamefont
  {Suweis}}, \bibinfo {author} {\bibfnamefont {F.}~\bibnamefont {Simini}},
  \bibinfo {author} {\bibfnamefont {J.~R.}\ \bibnamefont {Banavar}}, \ and\
  \bibinfo {author} {\bibfnamefont {A.}~\bibnamefont {Maritan}},\ }\href
  {http://dx.doi.org/10.1038/nature12438} {\bibfield  {journal} {\bibinfo
  {journal} {Nature}\ }\textbf {\bibinfo {volume} {500}},\ \bibinfo {pages}
  {449} (\bibinfo {year} {2013})}\BibitemShut {NoStop}%
\bibitem [{\citenamefont {Saavedra}\ \emph {et~al.}(2016)\citenamefont
  {Saavedra}, \citenamefont {Rohr}, \citenamefont {Olesen},\ and\ \citenamefont
  {Bascompte}}]{https://doi.org/10.1002/ece3.1930}%
  \BibitemOpen
  \bibfield  {author} {\bibinfo {author} {\bibfnamefont {S.}~\bibnamefont
  {Saavedra}}, \bibinfo {author} {\bibfnamefont {R.~P.}\ \bibnamefont {Rohr}},
  \bibinfo {author} {\bibfnamefont {J.~M.}\ \bibnamefont {Olesen}}, \ and\
  \bibinfo {author} {\bibfnamefont {J.}~\bibnamefont {Bascompte}},\ }\href
  {\doibase https://doi.org/10.1002/ece3.1930} {\bibfield  {journal} {\bibinfo
  {journal} {Ecol. Evol.}\ }\textbf {\bibinfo {volume} {6}},\ \bibinfo {pages}
  {997} (\bibinfo {year} {2016})}\BibitemShut {NoStop}%
\bibitem [{\citenamefont {Yan}\ \emph {et~al.}(2017)\citenamefont {Yan},
  \citenamefont {Martinez},\ and\ \citenamefont {Liu}}]{GanYan2017}%
  \BibitemOpen
  \bibfield  {author} {\bibinfo {author} {\bibfnamefont {G.}~\bibnamefont
  {Yan}}, \bibinfo {author} {\bibfnamefont {N.~D.}\ \bibnamefont {Martinez}}, \
  and\ \bibinfo {author} {\bibfnamefont {Y.~Y.}\ \bibnamefont {Liu}},\ }\href
  {\doibase 10.1098/rsif.2017.0189} {\bibfield  {journal} {\bibinfo  {journal}
  {J. Roy. Soc. Interface}\ }\textbf {\bibinfo {volume} {14}},\ \bibinfo
  {pages} {20170189} (\bibinfo {year} {2017})}\BibitemShut {NoStop}%
\bibitem [{\citenamefont {Maeng}\ \emph {et~al.}(2012)\citenamefont {Maeng},
  \citenamefont {Lee},\ and\ \citenamefont {Lee}}]{Maeng2012}%
  \BibitemOpen
  \bibfield  {author} {\bibinfo {author} {\bibfnamefont {S.~E.}\ \bibnamefont
  {Maeng}}, \bibinfo {author} {\bibfnamefont {J.~W.}\ \bibnamefont {Lee}}, \
  and\ \bibinfo {author} {\bibfnamefont {D.-S.}\ \bibnamefont {Lee}},\ }\href
  {\doibase 10.1103/PhysRevLett.108.108701} {\bibfield  {journal} {\bibinfo
  {journal} {Phys. Rev. Lett.}\ }\textbf {\bibinfo {volume} {108}},\ \bibinfo
  {pages} {108701} (\bibinfo {year} {2012})}\BibitemShut {NoStop}%
\bibitem [{\citenamefont {Pascual-Garc{\'\i}a}\ and\ \citenamefont
  {Bastolla}(2017)}]{Pascual-Garcia:2017wj}%
  \BibitemOpen
  \bibfield  {author} {\bibinfo {author} {\bibfnamefont {A.}~\bibnamefont
  {Pascual-Garc{\'\i}a}}\ and\ \bibinfo {author} {\bibfnamefont
  {U.}~\bibnamefont {Bastolla}},\ }\href {\doibase 10.1038/ncomms14326}
  {\bibfield  {journal} {\bibinfo  {journal} {Nat. Commun.}\ }\textbf {\bibinfo
  {volume} {8}},\ \bibinfo {pages} {14326} (\bibinfo {year}
  {2017})}\BibitemShut {NoStop}%
\bibitem [{\citenamefont {Barbier}\ \emph {et~al.}(2018)\citenamefont
  {Barbier}, \citenamefont {Arnoldi}, \citenamefont {Bunin},\ and\
  \citenamefont {Loreau}}]{barbier_generic_2018}%
  \BibitemOpen
  \bibfield  {author} {\bibinfo {author} {\bibfnamefont {M.}~\bibnamefont
  {Barbier}}, \bibinfo {author} {\bibfnamefont {J.-F.}\ \bibnamefont
  {Arnoldi}}, \bibinfo {author} {\bibfnamefont {G.}~\bibnamefont {Bunin}}, \
  and\ \bibinfo {author} {\bibfnamefont {M.}~\bibnamefont {Loreau}},\ }\href
  {\doibase 10.1073/pnas.1710352115} {\bibfield  {journal} {\bibinfo  {journal}
  {Proc. Natl. Acad. Sci. U.S.A.}\ }\textbf {\bibinfo {volume} {115}},\
  \bibinfo {pages} {2156} (\bibinfo {year} {2018})}\BibitemShut {NoStop}%
\bibitem [{\citenamefont {Gracia-L{\'a}zaro}\ \emph {et~al.}(2018)\citenamefont
  {Gracia-L{\'a}zaro}, \citenamefont {Hern{\'a}ndez}, \citenamefont
  {Borge-Holthoefer},\ and\ \citenamefont {Moreno}}]{Gracia-Lazaro:2018tg}%
  \BibitemOpen
  \bibfield  {author} {\bibinfo {author} {\bibfnamefont {C.}~\bibnamefont
  {Gracia-L{\'a}zaro}}, \bibinfo {author} {\bibfnamefont {L.}~\bibnamefont
  {Hern{\'a}ndez}}, \bibinfo {author} {\bibfnamefont {J.}~\bibnamefont
  {Borge-Holthoefer}}, \ and\ \bibinfo {author} {\bibfnamefont
  {Y.}~\bibnamefont {Moreno}},\ }\href {\doibase 10.1038/s41598-018-27498-8}
  {\bibfield  {journal} {\bibinfo  {journal} {Sci. Rep.}\ }\textbf {\bibinfo
  {volume} {8}},\ \bibinfo {pages} {9253} (\bibinfo {year} {2018})}\BibitemShut
  {NoStop}%
\bibitem [{\citenamefont {Wang}\ \emph {et~al.}(2021)\citenamefont {Wang},
  \citenamefont {Peron}, \citenamefont {Dubbeldam}, \citenamefont {K{\`e}fi},\
  and\ \citenamefont {Moreno}}]{wang2021interspecific}%
  \BibitemOpen
  \bibfield  {author} {\bibinfo {author} {\bibfnamefont {X.}~\bibnamefont
  {Wang}}, \bibinfo {author} {\bibfnamefont {T.}~\bibnamefont {Peron}},
  \bibinfo {author} {\bibfnamefont {J.~L.~A.}\ \bibnamefont {Dubbeldam}},
  \bibinfo {author} {\bibfnamefont {S.}~\bibnamefont {K{\`e}fi}}, \ and\
  \bibinfo {author} {\bibfnamefont {Y.}~\bibnamefont {Moreno}},\ }\href@noop {}
  {\bibfield  {journal} {\bibinfo  {journal} {arXiv:2102.02259}\ } (\bibinfo
  {year} {2021})}\BibitemShut {NoStop}%
\bibitem [{\citenamefont {Maeng}\ \emph {et~al.}(2019)\citenamefont {Maeng},
  \citenamefont {Lee},\ and\ \citenamefont {Lee}}]{Maeng2019}%
  \BibitemOpen
  \bibfield  {author} {\bibinfo {author} {\bibfnamefont {S.~E.}\ \bibnamefont
  {Maeng}}, \bibinfo {author} {\bibfnamefont {J.~W.}\ \bibnamefont {Lee}}, \
  and\ \bibinfo {author} {\bibfnamefont {D.-S.}\ \bibnamefont {Lee}},\ }\href
  {\doibase 10.1088/1742-5468/ab0549} {\bibfield  {journal} {\bibinfo
  {journal} {J. Stat. Mech.}\ }\textbf {\bibinfo {volume} {2019}},\ \bibinfo
  {pages} {033502} (\bibinfo {year} {2019})}\BibitemShut {NoStop}%
\bibitem [{\citenamefont {Cai}\ \emph {et~al.}(2020)\citenamefont {Cai},
  \citenamefont {Snyder}, \citenamefont {Hastings},\ and\ \citenamefont
  {D'Souza}}]{Cai:2020ta}%
  \BibitemOpen
  \bibfield  {author} {\bibinfo {author} {\bibfnamefont {W.}~\bibnamefont
  {Cai}}, \bibinfo {author} {\bibfnamefont {J.}~\bibnamefont {Snyder}},
  \bibinfo {author} {\bibfnamefont {A.}~\bibnamefont {Hastings}}, \ and\
  \bibinfo {author} {\bibfnamefont {R.~M.}\ \bibnamefont {D'Souza}},\ }\href
  {\doibase 10.1038/s41467-020-19154-5} {\bibfield  {journal} {\bibinfo
  {journal} {Nat. Commun.}\ }\textbf {\bibinfo {volume} {11}},\ \bibinfo
  {pages} {5470} (\bibinfo {year} {2020})}\BibitemShut {NoStop}%
\bibitem [{\citenamefont {James}\ \emph {et~al.}(2012)\citenamefont {James},
  \citenamefont {Pitchford},\ and\ \citenamefont {Plank}}]{James:2012aa}%
  \BibitemOpen
  \bibfield  {author} {\bibinfo {author} {\bibfnamefont {A.}~\bibnamefont
  {James}}, \bibinfo {author} {\bibfnamefont {J.~W.}\ \bibnamefont
  {Pitchford}}, \ and\ \bibinfo {author} {\bibfnamefont {M.~J.}\ \bibnamefont
  {Plank}},\ }\href {\doibase 10.1038/nature11214} {\bibfield  {journal}
  {\bibinfo  {journal} {Nature}\ }\textbf {\bibinfo {volume} {487}},\ \bibinfo
  {pages} {227} (\bibinfo {year} {2012})}\BibitemShut {NoStop}%
\bibitem [{\citenamefont {Saavedra}\ and\ \citenamefont
  {Stouffer}(2013)}]{Saavedra:2013aa}%
  \BibitemOpen
  \bibfield  {author} {\bibinfo {author} {\bibfnamefont {S.}~\bibnamefont
  {Saavedra}}\ and\ \bibinfo {author} {\bibfnamefont {D.~B.}\ \bibnamefont
  {Stouffer}},\ }\href {\doibase 10.1038/nature12380} {\bibfield  {journal}
  {\bibinfo  {journal} {Nature}\ }\textbf {\bibinfo {volume} {500}},\ \bibinfo
  {pages} {E1} (\bibinfo {year} {2013})}\BibitemShut {NoStop}%
\bibitem [{\citenamefont {Allesina}\ and\ \citenamefont
  {Tang}(2012)}]{Allesina2012}%
  \BibitemOpen
  \bibfield  {author} {\bibinfo {author} {\bibfnamefont {S.}~\bibnamefont
  {Allesina}}\ and\ \bibinfo {author} {\bibfnamefont {S.}~\bibnamefont
  {Tang}},\ }\href {\doibase 10.1038/nature10832} {\bibfield  {journal}
  {\bibinfo  {journal} {Nature}\ }\textbf {\bibinfo {volume} {483}},\ \bibinfo
  {pages} {205} (\bibinfo {year} {2012})}\BibitemShut {NoStop}%
\bibitem [{\citenamefont {Stone}(2020)}]{Stone2020}%
  \BibitemOpen
  \bibfield  {author} {\bibinfo {author} {\bibfnamefont {L.}~\bibnamefont
  {Stone}},\ }\href {\doibase 10.1038/s41467-020-16474-4} {\bibfield  {journal}
  {\bibinfo  {journal} {Nat. Commun.}\ }\textbf {\bibinfo {volume} {11}},\
  \bibinfo {pages} {2648} (\bibinfo {year} {2020})}\BibitemShut {NoStop}%
\bibitem [{web()}]{web_of_life}%
  \BibitemOpen
  \href@noop {} {}\bibinfo {note} {Http://www.web-of-life.es/}\BibitemShut
  {NoStop}%
\bibitem [{\citenamefont {Arroyo}\ \emph {et~al.}(1982)\citenamefont {Arroyo},
  \citenamefont {Primack},\ and\ \citenamefont
  {Armesto}}]{https://doi.org/10.1002/j.1537-2197.1982.tb13237.x}%
  \BibitemOpen
  \bibfield  {author} {\bibinfo {author} {\bibfnamefont {M.~T.~K.}\
  \bibnamefont {Arroyo}}, \bibinfo {author} {\bibfnamefont {R.}~\bibnamefont
  {Primack}}, \ and\ \bibinfo {author} {\bibfnamefont {J.}~\bibnamefont
  {Armesto}},\ }\href {\doibase
  https://doi.org/10.1002/j.1537-2197.1982.tb13237.x} {\bibfield  {journal}
  {\bibinfo  {journal} {Am. J. Bot.}\ }\textbf {\bibinfo {volume} {69}},\
  \bibinfo {pages} {82} (\bibinfo {year} {1982})}\BibitemShut {NoStop}%
\bibitem [{\citenamefont {Lee}\ \emph {et~al.}(2009)\citenamefont {Lee},
  \citenamefont {Ha}, \citenamefont {Jeong}, \citenamefont {Noh},\ and\
  \citenamefont {Park}}]{PhysRevE.80.051127}%
  \BibitemOpen
  \bibfield  {author} {\bibinfo {author} {\bibfnamefont {S.~H.}\ \bibnamefont
  {Lee}}, \bibinfo {author} {\bibfnamefont {M.}~\bibnamefont {Ha}}, \bibinfo
  {author} {\bibfnamefont {H.}~\bibnamefont {Jeong}}, \bibinfo {author}
  {\bibfnamefont {J.~D.}\ \bibnamefont {Noh}}, \ and\ \bibinfo {author}
  {\bibfnamefont {H.}~\bibnamefont {Park}},\ }\href {\doibase
  10.1103/PhysRevE.80.051127} {\bibfield  {journal} {\bibinfo  {journal} {Phys.
  Rev. E}\ }\textbf {\bibinfo {volume} {80}},\ \bibinfo {pages} {051127}
  (\bibinfo {year} {2009})}\BibitemShut {NoStop}%
\bibitem [{git()}]{githubcode}%
  \BibitemOpen
  \href@noop {} {}\bibinfo {note} {The matlab code is available at
  https://github.com/deoksunlee/Stability-and-selective-extinction-in-complex-mutualistic-networks}\BibitemShut
  {NoStop}%
\end{thebibliography}%
\end{document}